\documentclass[aps,prd,superscriptaddress,nofootinbib,amsmath,amsfonts,preprintnumbers,groupedaddress,showpacs,10pt,english]{revtex4-1}
\usepackage{amsmath}
\usepackage{amssymb}
\usepackage{babel}
\usepackage{wrapfig}
\usepackage{cancel}

\newcommand{\e}{\mathrm{e}}
\usepackage{relsize,exscale}
\makeatletter

\usepackage{array,multirow,graphicx}
\usepackage{dcolumn}
\usepackage{newlfont}
\usepackage{bm}
\usepackage[colorlinks,citecolor=blue,urlcolor=blue,linkcolor=blue]{hyperref}
\usepackage[figtopcap]{subfigure}
\usepackage{color}
\usepackage{verbatim}

\def\be{\begin{align}}
\def\ee{\end{align}}
\def\bea{\begin{eqnarray}}
\def\eea{\end{eqnarray}}
\def\bal{\begin{align}}
\def\eal{\end{align}}

\usepackage{scalerel}
\usepackage{tikz}
\usetikzlibrary{svg.path}
\definecolor{orcidlogocol}{HTML}{A6CE39}
\tikzset{
 orcidlogo/.pic={
 \fill[orcidlogocol] svg{M256,128c0,70.7-57.3,128-128,128C57.3,256,0,198.7,0,128C0,57.3,57.3,0,128,0C198.7,0,256,57.3,256,128z};
 \fill[white] svg{M86.3,186.2H70.9V79.1h15.4v48.4V186.2z}
 svg{M108.9,79.1h41.6c39.6,0,57,28.3,57,53.6c0,27.5-21.5,53.6-56.8,53.6h-41.8V79.1z M124.3,172.4h24.5c34.9,0,42.9-26.5,42.9-39.7c0-21.5-13.7-39.7-43.7-39.7h-23.7V172.4z}
 svg{M88.7,56.8c0,5.5-4.5,10.1-10.1,10.1c-5.6,0-10.1-4.6-10.1-10.1c0-5.6,4.5-10.1,10.1-10.1C84.2,46.7,88.7,51.3,88.7,56.8z};}}
\newcommand\orcid[1]{\href{https://orcid.org/#1}{\mbox{\scalerel*{
\begin{tikzpicture}[yscale=-1,transform shape]
\pic{orcidlogo};
\end{tikzpicture}
}{|}}}}
\graphicspath{{./}{Figs/}}
\begin{document}
\date{\today}


\title{  Four-scalar model and spherically symmetric solution in $f({\cal T})$  theory}

\author{G.~G.~L.~Nashed$^{1}$\orcid{0000-0001-5544-1119}}\email{nashed@bue.edu.eg}
\affiliation {Centre for Theoretical Physics, The British University in Egypt, P.O. Box
43, El Sherouk City, Cairo 11837, Egypt,\\
Centre for Space Research, North-West University, Potchefstroom, South Africa}
\author{A.~Eid$^{2}$\orcid{0000-0002-4189-0906}}\email{amaid@imamu.edu.sa}
\affiliation{Department of Physics, College of Science, Imam Mohammad Ibn Saud Islamic University (IMSIU), Riyadh, Kingdom of Saudi Arabia}
\begin{abstract}

In this work, we investigate the construction of spherically symmetric solutions within the framework of
modified teleparallel gravity, focusing in particular on $f({\cal T})$ theory, where ${\cal T}$ represents the torsion scalar.
Conventional mimetic gravity and its extension with two scalar fields are unable to reproduce general
spherically symmetric geometries in $f({\cal T})$ gravity, since consistency requires either a constant torsion scalar
or a linear form of $f({\cal T})$, which corresponds to the teleparallel equivalent of General Relativity (TEGR).
To overcome this restriction, we introduce a four-scalar field formulation that generalizes the two-scalar model
and enables the realization of arbitrary spherically symmetric spacetimes.
We further demonstrate that the model avoids ghost degrees of freedom by imposing appropriate constraints
through Lagrange multipliers. As an application of the formalism, we examine a specific spherically symmetric metric and determine the
corresponding four scalar fields, analyzing their properties.
In the presence of a quadratic extension of the action,
${\cal T} + \tfrac{\alpha}{2}{\cal T}^2$, our approach successfully reconstructs physically viable solutions
free from ghost instabilities. These findings highlight the potential of scalar-field-based constructions for generating consistent spacetime
geometries in modified gravity theories, thereby extending beyond the standard formulations.

\end{abstract}

\maketitle

\section{Introduction}
\label{introduction}

Albert Einstein introduced the theory of general relativity (GR) in 1915, establishing a profound link between matter and the curvature of spacetime via a system of nonlinear partial differential equations. This groundbreaking framework redefined gravity not as a force acting at a distance, but as the manifestation of spacetime curvature induced by matter. GR has been remarkably successful in explaining a wide range of gravitational phenomena, from Edwin Hubble's observation of cosmic redshift, which supports the notion of an expanding universe, to the imaging of black hole shadows and the detection of gravitational waves. Technological advancements, including satellites, high-precision detectors, telescopes, and numerical simulations, have transformed GR from a purely theoretical construct into a theory underpinned by robust empirical evidence. Contemporary observations and cosmological models indicate that approximately 95\% of the universe comprises a ``dark sector'' consisting of dark matter and dark energy. Understanding this component has become a central focus of modern astrophysics and cosmology, particularly in addressing galaxy rotation curves and the universe's accelerated expansion. Although classical GR remains a foundational theory in physics, the increasing precision of observational data underscores the need for its refinement or extension to accommodate new empirical findings.

In recent years, numerous alternative and modified theories of gravity have been proposed to investigate the dynamic behavior of the evolving Universe. These theories extend GR by altering the gravitational action, often through the replacement of the Ricci scalar  ${\cal R} $ with a more general function, such as  $f({\cal R})$, or by incorporating additional curvature invariants \cite{Nojiri:2010wj,Nashed:2019yto}. Other approaches involve coupling the Ricci scalar to scalar fields, introducing vector field contributions, or exploring gravity in higher-dimensional spacetimes. Among these frameworks, the
 $f({\cal T})$ theory, where  ${\cal T}$ represents the torsion scalar, has emerged as a compelling candidate. It offers promising avenues for explaining various cosmological observations and enhancing our theoretical comprehension of the Universe's evolution.

Instead of describing gravity through spacetime curvature using the Levi-Civita connection, one can adopt the Weitzenb\"ock connection, which is characterized by torsion rather than curvature. In this formulation, known as teleparallelism, torsion arises mainly from the first derivatives of the tetrad field, without involving second derivatives in the torsion tensor. Albert Einstein introduced this idea in 1928 as part of his search for a unified field theory \cite{Einstein:1929xrj, Einstein:1918btx, Unzicker:2005in}.

Teleparallelism closely resembles GR  but differs in geometric structure. It includes additional boundary terms in the action, total derivatives that do not affect the field equations. The theory uses the WeitzenbÃ¶ck spacetime, where the curvature tensor vanishes, metric compatibility is maintained, and spacetime is described through external translations. This connection to the group of universal coordinate transformations makes teleparallelism geometrically and physically appealing \cite{Bengochea:2010sg, HoffdaSilva:2009ht,Nashed:2005kn, Poplawski:2010kb, Wu:2010mn, Ao:2010mg, Wu:2010xk,ElHanafy:2014efn, Bengochea:2008gz, Lucas:2009nq, Ferraro:2008ey, Ferraro:2006jd}.

The teleparallel equivalent of general relativity (TEGR) formulates gravity using a Lagrangian linear in the torsion scalar, ${\cal T}$, with the tetrad (or vierbein) field serving as the fundamental dynamical variable instead of the metric. This tetrad acts as an orthonormal basis in the tangent space. While GR and TEGR differ geometrically, curvature vs. torsion, they are dynamically equivalent: every GR solution is also a TEGR solution.

Building on TEGR, the $f({\cal T})$ theories generalize the Lagrangian by replacing ${\cal T}$ with a general function $f({\cal T})$, analogous to how $f({\cal R})$ gravity extends GR. These models have attracted attention as promising alternatives for addressing cosmic acceleration \cite{Cardone:2012xq, Bengochea:2010sg, Karami:2010bys, Capozziello:2011hj, Bamba:2013jqa,Nashed:2021pah, Camera:2013bwa}, offering a torsion-based explanation distinct from dark energy.

Extensive research in $f({\cal T})$ gravity has explored a wide range of topics, including exact solutions and stellar structure models \cite{Gonzalez:2011dr, Nashed:2021pah, Wang:2011xf, Maurya:2023dhm, Nashed:2019zmy, Nashed:2018cth, Nashed:2011fz,Nashed:2015pga, Ferraro:2011ks, Nashed:2023srz, Capozziello:2012zj, Nashed:2013bfa}. In particular, spherically symmetric solutions play a crucial role in testing these theories. They help connect theoretical predictions with observations, such as planetary motion in the solar system, by modeling the gravitational field around point-like sources.

Although significant progress has been made in studying $f({\cal T})$ gravity across various physical contexts
~\cite{Ren:2022aeo, Nashed:2021pah, Chen:2019ftv, Cai:2019bdh, Nashed:2018cth, Capozziello:2019uvk, Nashed:2017fnd}, the theory still presents conceptual challenges when the functional form of $f({\cal T})$ is non-linear. In particular, the precise identification of physical degrees of freedom in such cases remains an open issue~\cite{Li:2011rn, Ferraro:2018axk}. Furthermore, it has been argued that for non-linear $f({\cal T})$ models, the existence of spherically symmetric geometries requires the torsion scalar ${\cal T}$ to be constant~\cite{Calza:2023hhi, DeBenedictis:2022sja}.

In this work, we propose an extended formulation that introduces four scalar fields, generalizing the two-scalar approach. Within this construction, we demonstrate the possibility of obtaining spherically symmetric configurations. To motivate this, we first revisit the two-scalar field model and highlight the need for a gravitational framework capable of reproducing general spherically symmetric spacetimes. In Section~\ref{fourscalar}, we develop the four-scalar extension and show that it can consistently reconstruct broad classes of geometries while avoiding ghost-like instabilities under appropriate conditions. Section~\ref{fQ} is devoted to an analysis of $f({\cal T})$ gravity, where we establish that, unlike the two-scalar case, the four-scalar model naturally accommodates spherically symmetric solutions. In Section~\ref{app}, we present an explicit example of our construction, computing the associated scalar fields as well as the relevant thermodynamic quantities. The main results and their implications are summarized in the final section.

The Latin indices, $i,j=1,2,3$, while the Greek labels $\mu,\nu=0,1,2,3$ are conventions. Additionally, we select $(-,+,+,+)$ as the metric's signature.

\section{Review of the Two-Scalar Model and its Limitations in $f(T)$ Gravity}
\label{twoscalar}
In Einstein's gravity, it has been shown in~\cite{Nojiri:2020blr} that a model involving two scalar fields coupled to the gravitational sector can reproduce arbitrary spherically symmetric spacetimes. Nevertheless, this formulation suffers from the appearance of ghost modes, which make the theory physically inconsistent. From the perspective of quantum theory, ghosts correspond to negative-norm states, which contradict the principles of the Copenhagen interpretation~\cite{Kugo:1979gm}.

Subsequent to the work of~\cite{Nojiri:2020blr}, further investigations~\cite{Nojiri:2023dvf, Nojiri:2023zlp, Elizalde:2023rds, Nojiri:2023ztz} demonstrated that the ghost problem can be circumvented by introducing constraints through Lagrange multipliers, thereby eliminating their propagation. These limits can be considered as a natural generalization of the mimetic condition first proposed in~\cite{Chamseddine:2013kea}\footnote{Mimetic gravity can be interpreted as a special realization of a conformal transformation in which the original and the transformed metrics become degenerate. In contrast, applying a non-singular conformal transformation increases the number of dynamical degrees of freedom, turning the longitudinal mode of gravity into a propagating one~\cite{Deruelle:2014zza, Domenech:2015tca, Firouzjahi:2018xob, Shen:2019nyp, Gorji:2020ten}. Typically, the conformal relation connects the physical metric ${\mathrm g}*{\alpha \beta}$ with an auxiliary metric $\bar{{\mathrm g}}*{\alpha \beta}$ and a scalar field $\psi$, in the form:}.

\begin{equation}
\label{trans1}
{\mathrm g}_{\alpha\beta}=\mp \left(\bar{{\mathrm g}}^{\mu \nu} \partial_\mu \psi \partial_\nu \psi \right) \bar{{\mathrm g}}_{\alpha\beta}\,.
\end{equation}
The transformation given in Eq.~(\ref{trans1}) leads to the following condition \cite{Gorji:2020ten}:
\begin{equation}
\label{trans2}
{\mathrm g}^{\alpha \beta}\partial_\alpha \psi \partial_\beta \psi= \mp 1\,.
\end{equation}
Thus, $\partial_\alpha \psi$ is timelike or spacelike depending on whether the negative or positive sign is chosen in Eq.(\ref{trans1}) or Eq.(\ref{trans2}), respectively. Here, we adopt the metric signature ${\mathrm g}_{\mu\nu} = \mathrm{diag}(-, +, +, +)$. The most well-known version of mimetic gravity corresponds to the negative sign, while the positive sign can be viewed as an extension of the theory.

A notable feature of the transformation in Eq.~(\ref{trans1}) is that it is non-invertible, meaning that the auxiliary metric $\bar{{\mathrm g}}_{\alpha \beta}$ cannot be expressed in terms of the physical metric ${\mathrm g}_{\alpha \beta}$ \cite{Deruelle:2014zza}. This non-invertibility introduces an extra degree of freedom, corresponding to the longitudinal mode of gravity.

Starting from the Einstein-Hilbert action written in terms of the physical metric $g_{\alpha \beta}$, applying the transformation in Eq.~(\ref{trans1}) yields an auxiliary metric $\bar{{\mathrm g}}_{\alpha \beta}$ and a dynamical scalar field $\psi$ \cite{Chaichian:2014qba}. 

In this section, we examine a general class of gravity theories minimally coupled to two scalar fields. As an extension of the model in~\cite{Nojiri:2020blr}, the action with two scalar fields is given, $\varphi$ and $\xi$:
\begin{align}
\label{I8B}
{\mathrm { \Gamma}}  =&\,{\mathrm {\Gamma} _\mathrm{gravity} + {\Gamma} _{\varphi\xi} + {\Gamma} _\mathrm{matter}} \, , \nonumber \\
{\mathrm{\Gamma} _{\varphi\xi}} =&\, {\mathrm\int d^4 x \sqrt{-{\mathrm g}} \left\{
 - \frac{1}{2} {\cal A} \mathrm{(\varphi,\xi) \partial_\mu \varphi \partial^\mu \varphi}
 - {\cal B} \mathrm{(\varphi,\xi) \partial_\mu \varphi \partial^\mu \xi
 - \frac{1}{2} {\cal C} (\varphi,\xi)  \partial_\mu \xi \partial^\mu \xi}
 - {\cal V}  (\varphi,\xi) \right\}}\, ,
\end{align}
where, ${\Gamma}\mathrm{gravity}$ represents the general action for the gravitational theory, while ${\Gamma}\mathrm{matter}$ describes the matter sector.  { Here, the functions $A(\varphi,\xi)$, ${\cal B}(\varphi,\xi)$, and ${\cal C}(\varphi,\xi)$ are arbitrary functions, all depending on the scalar fields $\varphi$ and $\xi$. These functions serve as the coefficients of the two scalar fields and vanish if the scalar fields are removed from the construction of the theory. Additionally, the potential ${\cal V}(\varphi,\xi)$ also depends on both $\varphi$ and $\xi$.}

Variation of Eq.~\eqref{I8B}  regarding ${\mathrm g}_{\mu\nu}$, we derive the  corresponding to the Einstein equation as,
\begin{align}
\label{gb4bD41}
&\,  \frac{1}{2} {\mathrm g}_{\mu\nu} \left\{{\mathrm
 - \frac{1}{2} {\cal A} (\varphi,\xi) \partial_\rho \varphi \partial^\rho \varphi
 - {\cal B} (\varphi,\xi) \partial_\rho \varphi \partial^\rho \xi
 - \frac{1}{2} {\cal C} (\varphi,\xi)  \partial_\rho \xi \partial^\rho \xi - {\cal V}  (\varphi,\xi)}\right\} \nonumber \\
&\,{\mathrm + \frac{1}{2} \left\{ {\cal A} (\varphi,\xi) \partial_\mu \varphi \partial_\nu \varphi
+ {\cal B} (\varphi,\xi) \left( \partial_\mu \varphi \partial_\nu \xi
+ \partial_\nu \varphi \partial_\mu \xi \right)
+ {\cal C} (\varphi,\xi)  \partial_\mu \xi \partial_\nu \xi \right\}
+ \frac{1}{2} \left( T_\mathrm{matter} \right)_{\mu\nu}}\nonumber\\
&\,\mbox{where} \quad  \left( T_\mathrm{matter} \right)_{\mu\nu}\quad \mbox{defines the energy-momentum tensor of matter as} \left( T_\mathrm{matter} \right)_{\mu\nu}\equiv\frac{2}{\sqrt{-g}} \frac{\delta {\cal S} _\mathrm{matter}}{\delta {\mathrm g}_{\mu\nu}} \, .
\end{align}
 Variation of ${\mathrm { \Gamma}} _\mathrm{gravity}$ of the gravity sector  yields:
\begin{align}
\label{mathcalG}
\mathcal{E}^{\mu\nu} \equiv - \frac{1}{\sqrt{-{\mathrm g}}} \frac{\delta {\mathrm { \Gamma}}_\mathrm{gravity}}{\delta {\mathrm g}_{\mu\nu}}\, .
\end{align}
Using Eqs.~(\ref{gb4bD41}) and  (\ref{mathcalG}) we get  the field equations of the action \eqref{I8B}, ${\mathrm { \Gamma}}$, in the form:
\begin{align}
\label{gb4bD4}
\mathcal{E}_{\mu\nu}=  &\, \frac{1}{2\kappa^2}\left(- R_{\mu\nu} + \frac{1}{2} g_{\mu\nu} R\right) \nonumber \\
&\, + \frac{1}{2} {\mathrm g}_{\mu\nu} \left\{
 - \frac{1}{2} A (\varphi,\xi) \partial_\rho \varphi \partial^\rho \varphi
 - B (\varphi,\xi) \partial_\rho \varphi \partial^\rho \xi
 - \frac{1}{2} C (\varphi,\xi) \partial_\rho \xi \partial^\rho \xi - V (\varphi,\xi)\right\} \nonumber \\
&\, + \frac{1}{2} \left\{ A (\varphi,\xi) \partial_\mu \varphi \partial_\nu \varphi
+ B (\varphi,\xi) \left( \partial_\mu \varphi \partial_\nu \xi
+ \partial_\nu \varphi \partial_\mu \xi \right)
+ C (\varphi,\xi) \partial_\mu \xi \partial_\nu \xi \right\} \nonumber \\
&\, - 2 \left[ \nabla_\mu \nabla_\nu \xi(\varphi,\xi)\right]R
+ 2 g_{\mu\nu} \left[ \nabla^2 \xi(\varphi,\xi)\right]R
+ 4 \left[ \nabla_\rho \nabla_\mu \xi(\varphi,\xi)\right]R_\nu^{\ \rho}
+ 4 \left[ \nabla_\rho \nabla_\nu \xi(\varphi,\xi)\right]R_\mu^{\ \rho} \nonumber \\
&\, - 4 \left[ \nabla^2 \xi(\varphi,\xi) \right]R_{\mu\nu}
 - 4g_{\mu\nu} \left[ \nabla_\rho \nabla_\sigma \xi(\varphi,\xi) \right] R^{\rho\sigma}
+ 4 \left[\nabla^\rho \nabla^\sigma \xi(\varphi,\xi) \right] R_{\mu\rho\nu\sigma}
+ \frac{1}{2} T_{\mathrm{matter}\, \mu\nu}\equiv0 \, ,
\end{align}
Additionally, variation  of the action \eqref{I8B}, ${\Gamma} _{\varphi\xi}$, w.r.t. $\varphi$ and $\xi$, give:
\begin{align}
\label{I10}
0 =&\, \frac{1}{2} {\cal A} _\varphi \partial_\mu \varphi \partial^\mu \varphi
+ {\cal A}  \nabla^\mu \partial_\mu \varphi + {\cal A} _\xi \partial_\mu \varphi \partial^\mu \xi
+ \left( {\cal B} _\xi - \frac{1}{2} {\cal C} _\varphi \right)\partial_\mu \xi \partial^\mu \xi
+ {\cal B}  \nabla^\mu \partial_\mu \xi - {\cal V} _\varphi \, , \nonumber \\
0 =&\, \left( - \frac{1}{2} {\cal A} _\xi + {\cal B} _\varphi \right) \partial_\mu \varphi \partial^\mu \varphi
+ {\cal B}  \nabla^\mu \partial_\mu \varphi
+ \frac{1}{2} {\cal C} _\xi \partial_\mu \xi \partial^\mu \xi + {\cal C}  \nabla^\mu \partial_\mu \xi
+ {\cal C} _\varphi \partial_\mu \varphi \partial^\mu \xi - {\cal V} _\xi \, ,
\end{align}
where, ${\cal A} _\varphi\equiv\frac{\partial {\cal A} (\varphi,\xi)}{\partial \varphi}$, etc.
If the equation $\nabla^\mu \mathcal{G}_{\mu\nu}=0$ is verified, then the Bianchi identities are recovered.

Spherical and time-varying spacetime is described by the following  metric in its general form \cite{Nojiri:2020blr},

\begin{align}
\label{GBiv}
ds^2{\mathcal  = - \e^{2\mu (t,r)} dt^2 + \e^{2\nu (t,r)} dr^2 +d\Omega^2\,, \quad \mbox{where}\quad d\Omega^2= r^2 \left( d\vartheta^2 + \sin^2\vartheta d\varphi^2 \right)}\, ,
\end{align}
where $\mu (t,r)$ and $\nu (t,r)$ are two arbitrary functions of $r$ and $t$.
It can be assumed without any loss of generality that
\begin{align}
\label{TSBH1}
\varphi=t\, , \qquad \qquad \xi=r\,, \quad \mbox{see \cite{Nojiri:2020blr, Nojiri:2023dvf, Nojiri:2023zlp, Elizalde:2023rds, Nojiri:2023ztz} for the detailed arguments}.
\end{align}

As mentioned in \cite{Nojiri:2020blr, Nojiri:2023dvf, Nojiri:2023zlp, Elizalde:2023rds, Nojiri:2023ztz}, the functions ${\cal A}$ and/or ${\cal C}$ often turn out to be negative, which leads to $\varphi$ and/or $\xi$ becoming negative, implying the presence of ghosts.

Through the introduction of the Lagrange multiplier $\lambda_\varphi$ and $\lambda_\xi$, we can impose constraints to eliminate the ghosts from the action ${\Gamma}$. Therefore, we modify the action ${\Gamma}_{\varphi\xi}$ in Eq.~(\ref{I8B}) as ${\Gamma} \to {\Gamma} + {\Gamma}\lambda$, where
\begin{align}
\label{lambda1}
{\mathrm {\Gamma} _\lambda = \int d^4 x \sqrt{-{\mathrm g}} \left[ \lambda_\varphi \left( \e^{2\mu(t=\varphi, r=\xi)} \partial_\mu \varphi \partial^\mu \varphi + 1 \right)
+ \lambda_\xi \left( \e^{2\nu(t=\varphi, r=\xi)}\partial_\mu \xi \partial^\mu \xi - 1 \right) \right]} \, .
\end{align}
Variation of  ${\mathrm {\Gamma}} _\lambda$ w.r.t. $\lambda_\varphi$ and $\lambda_\xi$ yields  (\ref{TSBH1}):
\begin{align}
\label{lambda2}
0 ={\mathrm \e^{2\mu(t=\varphi, r=\xi)} \partial_\mu \varphi \partial^\mu \varphi + 1} \, , \quad
0 ={\mathrm \e^{2\nu(t=\varphi, r=\xi)} \partial_\mu \xi \partial^\mu \xi - 1} \, .
\end{align}
Because of the limits presented in Eq.~(\ref{lambda2}), $\varphi$ and $\xi$ are no longer dynamical, and their fluctuations in Eq.(\ref{TSBH1}) no longer propagate. The changes are described as follows:
\begin{align}
\label{pert1}
{\mathrm\varphi=t + \delta \varphi} \, , \quad {\mathrm \xi=r + \delta \xi}\, .
\end{align}
Utilizing Eq.~(\ref{lambda2}), yields:
\begin{align}
\label{pert2}
{\mathrm \partial_t \left( \e^{2\mu({ t=\varphi, r=\xi})} \delta \varphi \right) = \partial_r \left( \e^{2\nu({ t=\varphi, r=\xi})} \delta \xi \right) = 0}\, .
\end{align}
According to Eq.~(\ref{pert2}), when we impose the initial condition $\delta\varphi = 0$ (since $\varphi$ corresponds to the time coordinate) and the boundary condition $\delta\xi \to 0$ as $r \to \infty$ (since $\xi$ corresponds to the radial coordinate), both $\delta\varphi$ and $\delta\xi$ vanish. Specifically, $\delta\varphi = 0$ at the initial surface and $\delta\xi = 0$ at the boundary surface. This indicates that $\varphi$ and $\xi$ are frozen, representing non-dynamical degrees of freedom.


Now, we are going to derive a solution realizing the ansatzs of the metric, i.e.,    $\e^{2\mu(t,r)}$ and $\e^{2\nu(t,r)}$ presented in (\ref{GBiv}).

Using Eq. \eqref{GBiv} in Eq. \eqref{I10} we get:
\begin{align}
\label{TSBH2B}
{\mathrm 2\mathcal{E}_{tt}} =&\, {\mathrm- \e^{2\mu} \left( - \frac{{\cal A} }{2} \e^{-2\mu} - \frac{{\cal C} }{2} \e^{-2\nu} - {\cal V}  \right) + \e^{2\mu} \rho} \, , \nonumber \\
{\mathrm2\mathcal{E}_{rr}} =&\,{\mathrm \e^{2\nu} \left( \frac{{\cal A} }{2} \e^{-2\mu} + \frac{{\cal C} }{2} \e^{-2\nu} - {\cal V}  \right) + \e^{2\nu} p} \, , \nonumber \\
{\mathrm 2\mathcal{E}_{\vartheta \vartheta } = \frac{2\mathcal{G}_{\varphi\varphi}}{\sin^2 \vartheta}}
=&\, {\mathrm r^2 \left( \frac{{\cal A} }{2} \e^{-2\mu} - \frac{{\cal C} }{2} \e^{-2\nu} - {\cal V}  \right) + r^2 p} \, , \nonumber \\
{\mathrm 2\mathcal{E}_{tr} = 2\mathcal{E}_{rt} }=&\,{\mathrm {\cal B}}  \, , \nonumber \\
0=&\, {\mathrm \mathcal{E}_{t\vartheta } = \mathcal{E}_{\vartheta  t} = \mathcal{E}_{r\vartheta } = \mathcal{G}_{\vartheta  r}
= \mathcal{E}_{t\varphi } = \mathcal{E}_{\varphi  t} = \mathcal{E}_{r\varphi } = \mathcal{E}_{\varphi  r}}\, .
\end{align}
In this study, we denote the matter energy density and pressure as $\rho$ and $p$, respectively. We assume that the matter stress-energy tensor $\left( T_\mathrm{matter} \right)_{\mu\nu}$ takes the form of a perfect fluid:
\begin{align}
\label{FRk2}
\left({\mathrm T_\mathrm{matter}}\right)_{tt}=\rho=-{\mathrm g}_{tt}\rho\ ,\quad \left( {\mathrm T_\mathrm{matter}}\right)_{ij} ={\mathrm p{\mathrm g}_{ij}}\, .
\end{align}
{ For Einstein's theory of gravity, we discover that  $\mathcal{E}_{t\vartheta }=\mathcal{E}_{\vartheta  t}$,
$\mathcal{E}_{r\vartheta }=\mathcal{E}_{\vartheta  r}$,
$\mathcal{E}_{t\varphi }=\mathcal{E}_{\varphi  t}$, and $\mathcal{E}_{r\varphi }=\mathcal{E}_{\varphi  r}$ are vanishing.
 Later, we will observe that in $f({\cal T})$ gravity, $\mathcal{E}_{r\vartheta} = \mathcal{E}_{\vartheta r}$ does not vanish. The condition $\mathcal{E}_{r\vartheta} = \mathcal{E}_{\vartheta r} = 0$ imposes significant constraints on the model, causing it to reduce to the TEGR case.

If $\mathcal{E}_{t\vartheta }=\mathcal{E}_{\vartheta  t}$, $\mathcal{E}_{r\vartheta }=\mathcal{E}_{\vartheta  r}$,
$\mathcal{E}_{t\varphi }=\mathcal{E}_{\varphi  t}$, and $\mathcal{E}_{r\varphi }=\mathcal{E}_{\varphi  r}$  are vanishing  then we can solve Eqs.~(\ref{TSBH2B}) w.r.t. ${\cal A} $, ${\cal B} $, ${\cal C} $, and ${\cal V} $ in the following manner:
\begin{align}
\label{ABCV}
&{\mathrm  {\cal A} = 2\left( \mathcal{E}_{tt} + \frac{\e^{2\mu}}{r^2} \mathcal{E}_{\vartheta \vartheta } \right) - \e^{2\mu} \left( \rho + p \right) }\, , \quad
{\mathrm{\cal B} = 2 \mathcal{E}_{tr}} \, , \quad
{\mathrm {\cal C} = 2\left( \mathcal{E}_{rr} - \frac{\e^{2\nu}}{r^2} \mathcal{E}_{\vartheta \vartheta } \right)} \, , \nonumber \\
& {\mathrm {\cal V} = \left( \e^{-2\mu} \mathcal{E}_{tt} - \e^{-2\nu} \mathcal{E}_{rr} \right) - \frac{1}{2} \left( \rho - p \right)} \, .
\end{align}
The spacetime defined by Eq.(\ref{GBiv}) can be realized through a model that replaces the coordinates $(t, r)$ in Eq.(\ref{ABCV}) with the two scalar fields $(\varphi, \xi)$. Before proceeding to the next section, we will summarize the results of this section. As shown in Eq.~(\ref{ABCV}), within the framework of Einstein's theory with two scalar fields, one can recover a spherically symmetric solution. But is this procedure valid within $f({\cal T})$ theory? We will discuss this in the next section.}

\section{$f({\cal T})$ theory}
\label{fQ}
While General Relativity (GR) interprets gravity as a manifestation of spacetime curvature,
alternative formulations can be developed by employing a different geometric quantity---the torsion tensor.
An important example is $f({\cal T})$ gravity, where the gravitational action is expressed in terms of a torsion scalar ${\cal T}$,
constructed from contractions of the torsion tensor. The teleparallel equivalent of General Relativity (TEGR) is recovered
when $f({\cal T})$ is taken to be a linear function of ${\cal T}$.
Despite its potential, the systematic development of consistent and observationally viable $f({\cal T})$ models
is still relatively limited.

In this section, we provide a brief overview of $f({\cal T})$ gravity~\cite{Bengochea:2008gz,Capozziello:2012zj,Awad:2017tyz}
and further discuss an extension of the theory that incorporates additional scalar fields.

\subsection{Brief review}
In GR, the metric tensor, defined on a pseudo-Riemannian manifold, acts as the fundamental dynamical variable. In contrast, within the framework of
${\cal T}$ gravity, the tetrad field serves as the dynamical quantity, establishing an orthonormal basis in the tangent space. In this context, Greek indices
$(\mu, \nu, ...)$ denote spacetime coordinates, while Latin indices $(i, j, ...)$ represent coordinates in the tangent space. The vectors and covectors in the tangent space are given by $h_i=\partial_i$ and $h^i = dx^i$, respectively. Similarly, the basis vectors and covectors in spacetime are denoted by
$h_\mu=\partial_\mu$ and $h^\mu = dx^\mu$. The relationship between these bases can be defined as:
\textrm{
\begin{equation}
    \begin{array}{ccc}
    \partial _i={h_i}^{\mu }\partial _{\mu } & \quad \mbox{and} \quad &  {dx}^i={h^i}_{\mu } {dx}^{\mu } ,
\end{array}
\label{E01}
\end{equation}
}
and conversely
\textrm{
\begin{equation}
    \begin{array}{ccc}
   \partial _{\mu }={h^i}_{\mu }\partial _i & \quad \mbox{and} \quad &  {dx}^{\mu }={h_i}^{\mu }{dx}^i.
\end{array}
\label{E02}
\end{equation}
}
The tetrad field ${h^i}_{\mu }$ and its inverse ${h_i}^{\mu }$ satisfy the relation ${h_j}^{\mu } {h^i}_{\mu }=\delta_j^i$ and ${h^i}^{\nu } {h_i}_{\mu }={\delta_\nu}^\mu$.  The connection between the tetrad field and the spacetime metric is given by:
\textrm{
\begin{equation}
{\mathrm g}_{\mu \nu }=\eta _{_{ij}}{h^i}_{\mu }{h^j}_{\nu } ,
    \label{E03}
\end{equation}
}
with $\eta_{_{ij}}=diag[-1,1,1,1]$ and $\sqrt{-{\mathrm g}}=det[{h^i}_{\mu }]=h$. Using Eq. (\ref{E01}), we transformed the tetrad field equations into the form:
\textrm{
\begin{equation}
{ds}^2={\mathrm g}_{\mu \nu }{dx}^{\mu }{dx}^{\nu }=\eta _{{ij}}{h^i}_{\mu }{h^j}_{\nu }{dx}^{\mu }{dx}^{\nu } .
    \label{E04}
\end{equation}
}
The Weitzenb\"ock connection  in $f({\cal T})$ gravity theory was introduced by \cite{Wr}:
\textrm{
\begin{equation}
{\Gamma^\zeta}_{\mu \nu}={h_i}^{\zeta}\partial _{\nu }{h^i}_{\mu }=-{h_i}^{\mu }\partial _{\nu }{h_i}^{\zeta } .
    \label{E05}
\end{equation}
}
The torsion tensor of  Weitzenb\"ock geometry is defined as follows:
\textrm{
\begin{equation}
{{\cal T}^\zeta }_{\mu \nu }={\Gamma^\zeta }_{\mu \nu }-{\Gamma^\zeta} _{\nu \mu }={h_i}^{\zeta }\left(\partial _{\mu }{h^i}_{\nu }-\partial _{\nu }{h^i}_{\mu }\right).
    \label{E06}
\end{equation}
}
Hence, the  contorsion, the superpotential, and the torsion scalar can be evaluated as follows:
\textrm{
\begin{align}\label{E07}
&{K_\zeta }^{\mu \nu }=-\frac{1}{2}\left({{\cal T}_\zeta }^{\mu \nu }-{{\cal T}_\zeta }^{\nu \mu }-{{\cal T}_\zeta }^{\mu \nu }\right) , \nonumber\\
&{S_\zeta }^{\mu \nu }=\frac{1}{2}\left({K_\zeta }^{\mu \nu }-{\delta_\zeta }^{\mu }{{\cal T}_\gamma }^{\gamma \nu }-{\delta_\zeta }^{\nu }{{\cal T}_\gamma }^{\gamma \mu }\right),    \nonumber\\
&{\cal T}={{\cal T}^\zeta }_{\mu \nu }{S_\zeta }^{\mu \nu } .
\end{align}
}
Like in $f(R)$ theory, in $f({\cal T})$ gravitational theory, the revised gravitational action can be expressed using the torsion scalar with the constants $(G=c=1)$.
\textrm{
\begin{equation}
S\left[{h^i}_{\mu },\phi_A\right]=\int d^4x\, h \left[\frac{1}{16 \pi }f({\cal T})+\mathcal{L}_{ {matter}}\left(\phi_A\right)\right] .
    \label{E10}
\end{equation}
}
Here $f({\cal T})$ represents a function related to the torsion scalar, while $\mathcal{L}_{ {matter}}$ signifies the Lagrangian density of the matter field.
The field equation for $f({\cal T})$ gravity are derived  from Eq.~(\ref{E10}) \cite{Bengochea:2008gz,Capozziello:2012zj,Awad:2017tyz} as follows:
\textrm{
\begin{equation}
  {I_\mu }^{\nu } \equiv  {S_\mu }^{\nu \gamma }\partial _{\gamma }{\cal T} f_{{{\cal T}{\cal T}}}+\left[h^{-1}{h^i}_{\mu } \partial _{\gamma }\left(h {h_i}^{\zeta } {S_\zeta }^{\nu \gamma }\right)+ {{\cal T}_{\theta \mu }}^{\zeta } {S_\zeta }^{\nu \theta }\right] f_{\cal T} + \frac{1}{4}{\delta _\mu }^{\nu }f-4 \pi  {T _\mu }^{\nu } .
\label{E11}
\end{equation}
}
Here ${{\cal T}_\mu }^{\nu } $ represents the energy-moment tensor of the matter, whihle $f_{\cal T}$ and $f_{{\cal T}{\cal T}}$ denote the first and second derivatives of
 $f({\cal T})$ with respect to ${\cal T}$, respectively.

\subsection{Spherically symmetric spacetime solution}\label{f(T)}

Assuming a general form of a spherically symmetric spacetime having a stationary spacetime, where the ansatzs in Eq.~ (\ref{GBiv}) depend on $\mu = \mu(r)$ and $\nu = \nu(r)$, we obtain
\textrm{
\begin{align}
\label{Qstatic}
{\cal T} = -{\frac { 2\left(1-2\,{e^{-\lambda
 }}+{e^{-2\lambda}}+2\,{e^{-2\,\lambda}}\nu' r-2\,{e^{-\lambda}}\nu'r \right)}{{r}^{2}}}\, .
\end{align}
}
We now examine the component $\mathcal{E}_{r\vartheta}$,
\textrm{
\begin{align}
\label{rtheta}
\mathcal{E}_{r\vartheta} = \frac{1}{2\tan\vartheta} \frac{d\left( f'({\cal T})\right)}{d{\cal T}}\frac{d{\cal T}}{dr}= \frac{1}{2\tan\vartheta} \frac{d\left( f'({\cal T})\right)}{dr} \,,
\end{align}
where  $f'$  denotes the derivative for the radial coordinate  $r$,  i.e.,  $f'=\frac{df}{dr}$
}
By imposing the condition  $\mathcal{E}_{r\theta} = 0$, we deduce that either $f''({\cal T})=0$  or the torsion scalar,  ${\cal T}$,  remains constant.
If $f''({\cal T}) = 0$, then $f({\cal T})$ becomes a linear function of ${\cal T}$, which coincides with GR. On the other hand, the constraint  ${\cal T}=const.$  imposes a significant limitation on the spacetime. To illustrate this, let us consider  Schwarzschild geometry, where the scalar torsion ${\cal T}$ is not constant; it is expressed as ${\cal T} = -\frac{2}{r^2}$. Consequently, within the context of $f({\cal T})$ theory, Schwarzschild spacetime is not a solution \cite{Awad:2017tyz}. Furthermore,  as explained in Sec.~\ref{twoscalar}, it is not possible to construct a general spherically symmetric spacetime within $f({\cal T})$ gravity. However, within the context of the four-scalar framework introduced in Sec.~\ref{fourscalar} it is possible to construct a general form of a spacetime that possesses spherical symmetry.

\section{Four-Scalar Field Model in $f({\cal T})$}
\label{fourscalar}

In certain theories, such as $f({\cal T})$  gravity, deriving spherically symmetric solutions proves challenging. As discussed in the previous section, the following components are non-zero:  \begin{align} \mathcal{E}_{t\vartheta }=\mathcal{E}_{\vartheta t},\quad  \mathcal{E}_{r\vartheta }=\mathcal{E}_{\theta r}, \quad \mbox{and} \quad \mathcal{E}_{t\varphi }=\mathcal{E}_{\varphi t}.\end{align}  This difficulty remains even when employing the two-scalar model introduced in Sec.~\ref{twoscalar}.
To overcome the restrictions associated with spherical symmetry in such frameworks,
we generalize the two-scalar setup to a four-scalar formulation.
This extended model allows for the realization of a  spherically symmetric spacetime,
thereby resolving the limitations inherent in the original construction.

Utilizing the four-scalar field model, we now examine the following action, in which $\varphi^{\left(\rho\right)}\left(\rho=0,1,2,3\right) $:
\begin{align}
\label{acg1}
{\mathrm {\Gamma}}  =&\,{\mathrm {\Gamma} _\mathrm{gravity} + {\Gamma} _\varphi + {\Gamma} _\lambda} \, , \nonumber \\
{\mathrm {\Gamma} _\varphi} \equiv&\, \int {\mathrm  d^4x \sqrt{-{\mathrm g}} \left\{ \frac{1}{2}{\mathit   \sum_{\rho,\sigma=0,1,2,3} \Phi _{\left(\rho\sigma\right)} \left(\varphi^{\left(\tau\right)}\right)
{\mathrm g}^{\mu\nu} \partial_\mu \varphi^{\left(\rho\right)} \partial_\nu \varphi^{\left(\sigma\right)} - {\cal V} \left( \varphi^{\left(\rho\right)} \right)} \right\}}\, , \nonumber \\
{\mathit {\Gamma} _\lambda} \equiv&\, \int {\mathrm  d^4x \sqrt{-{\mathrm g}} \sum_{\rho=0,1,2,3} \lambda^{(\rho)} \left( \frac{1}{{\mathrm g}^{\rho\rho}\left( x^\sigma = \varphi^{\left( \sigma \right)} \right)} {\mathrm g}^{\mu\nu} \left( x^\sigma \right)
\partial_\mu \varphi^{\left(\rho\right)} \partial_\nu \varphi^{\left(\rho\right)} - 1\right)} \, .
\end{align}
The quantities  ${\cal V} \left( \varphi^{\left(\rho\right)} \right)$ and $\Phi _{\left(\rho\sigma\right)} \left(\varphi^{\left(\tau\right)}\right)$   depend on  $\varphi^{\left(\rho\right)}$. Notably, the indices $\rho\sigma$ in $\Phi _{(\rho\sigma)}$,  and similar expressions, do not correspond to the tensor's indices. We employ bracket notation to distinguish between them  $\varphi^{(\rho)}$ with $x^\rho$, we use Greek indices. Their inclusion, while straightforward, leads to more complex expressions; thus, for simplicity, we omit the contributions from matter.
The component $\frac{1}{{\mathrm g}^{\rho\rho}\left( x^\sigma = \varphi^{\left( \sigma \right)} \right)}$ in ${\Gamma} _\lambda$ could potentially disrupt  diffeomorphism invariance. However, it is important to acknowledge that this factor derives from the scalar fields, not the metric, making it compatible with diffeomorphism invariance.

Within the context of ${\Gamma} _\lambda$, the $\lambda^{(\rho)}$ fields act as Lagrange multipliers that impose constraints similar to those in equation (\ref{lambda2}),
\textit{
\begin{align}
\label{cnstrnt1}
0 = \frac{1}{{\mathrm g}^{\rho\rho}\left( x^\sigma = \varphi^{\left( \sigma \right)} \right)} {\mathrm g}^{\mu\nu} \left( x^\sigma \right)
\partial_\mu \varphi^{\left(\rho\right)} \partial_\nu \varphi^{\left(\rho\right)} - 1 \, ,
\end{align}
}
that ensure the elimination of ghosts.

Taking the derivative of  Eq.~(\ref{acg1}) w.r.t.  ${\mathrm g}_{\mu\nu}$ yields:
\textit{
\begin{align}
\label{Eqs1}
\mathcal{E}_{\mu\nu} =&\, \frac{1}{2} {\mathrm g}_{\mu\nu} \left\{ \frac{1}{2} \sum_{\rho,\sigma=0,1,2,3} \Phi _{\left(\rho\sigma\right)} \left(\varphi^{\left(\tau\right)}\right)
{\mathrm g}^{\xi\eta} \partial_\xi \varphi^{\left(\rho\right)} \partial_\eta \varphi^{\left(\sigma\right)} - {\cal V} \left( \varphi^{\left(\rho\right)} \right) \right\}
 - \frac{1}{2} \sum_{\rho,\sigma=0,1,2,3} \Phi _{\left(\rho\sigma\right)} \left(\varphi^{\left(\tau\right)}\right)
\partial_\mu \varphi^{\left(\rho\right)} \partial_\nu \varphi^{\left(\sigma\right)} \nonumber \\
&\, + \frac{1}{2} {\mathrm g}_{\mu\nu} \sum_{\rho=0,1,2,3} \lambda^{(\rho)} \left(
\frac{1}{{\mathrm g}^{\rho\rho}\left( x^\sigma = \varphi^{\left( \sigma \right)} \right)} {\mathrm g}^{\xi\eta} \left( x^\sigma \right)
\partial_\xi \varphi^{\left(\rho\right)} \partial_\eta \varphi^{\left(\rho\right)} - 1\right)
 - \sum_{\rho=0,1,2,3} \frac{\lambda^{(\rho)}}{{\mathrm g}^{\rho\rho}\left( x^\sigma = \varphi^{\left( \sigma \right)} \right)}
\partial_\mu \varphi^{\left(\rho\right)} \partial_\nu \varphi^{\left(\rho\right)} \nonumber \\
=&\, \frac{1}{2} {\mathrm g}_{\mu\nu} \left\{ \frac{1}{2} \sum_{\rho,\sigma=0,1,2,3} \Phi_{\left(\rho\sigma\right)} \left(\varphi^{\left(\tau\right)}\right)
{\mathrm g}^{\xi\eta} \partial_\xi \varphi^{\left(\rho\right)} \partial_\eta \varphi^{\left(\sigma\right)} - {\cal V} \left( \varphi^{\left(\rho\right)} \right) \right\}
 - \frac{1}{2} \sum_{\rho,\sigma=0,1,2,3} \Phi _{\left(\rho\sigma\right)} \left(\varphi^{\left(\tau\right)}\right)
\partial_\mu \varphi^{\left(\rho\right)} \partial_\nu \varphi^{\left(\sigma\right)} \nonumber \\
&\, - \sum_{\rho=0,1,2,3} \frac{\lambda^{(\rho)}}{{\mathrm g}^{\rho\rho}\left( x^\sigma = \varphi^{\left( \sigma \right)} \right)}
\partial_\mu \varphi^{\left(\rho\right)} \partial_\nu \varphi^{\left(\rho\right)} \, ,
\end{align}
}
where Eq.~(\ref{cnstrnt1}) is used.
If Eq.~(\ref{Eqs1}) is multiplied by ${\mathrm g}^{\mu\nu}$ we get:
\textit{
\begin{align}
\label{Eqs2}
{\mathrm g}^{\mu\nu} \mathcal{E}_{\mu\nu} =&\,
\frac{1}{2} \sum_{\rho,\sigma=0,1,2,3} \Phi _{\left(\rho\sigma\right)} \left(\varphi^{\left(\tau\right)}\right)
{\mathrm g}^{\xi\eta} \partial_\xi \varphi^{\left(\rho\right)} \partial_\eta \varphi^{\left(\sigma\right)} - 2 {\cal V} \left( \varphi^{\left(\rho\right)} \right)
 - \sum_{\rho=0,1,2,3} \lambda^{(\rho)} \, ,
\end{align}
}
where Eq.~(\ref{cnstrnt1}) is used.
Through the use of Eq.~ (\ref{Eqs2}) in  Eq.~(\ref{Eqs1}), we get
\textit{
\begin{align}
\label{Eqs3}
\sum_{\rho,\sigma=0,1,2,3} \Phi _{\left(\rho\sigma\right)} \left(\varphi^{\left(\tau\right)}\right) \partial_\mu \varphi^{\left(\rho\right)} \partial_\nu \varphi^{\left(\sigma\right)}
=&\, - 2 \mathcal{E}_{\mu\nu}
+ {\mathrm g}_{\mu\nu} \left\{ {\cal V} \left( \varphi^{\left(\rho\right)} \right) + \sum_{\rho=0,1,2,3} \lambda^{(\rho)} + {\mathrm g}^{\rho\sigma} \mathcal{E}_{\rho\sigma} \right\} \nonumber \\
&\, - 2 \sum_{\rho=0,1,2,3} \frac{\lambda^{(\rho)}}{{\mathrm g}^{\rho\rho}\left( x^\sigma = \varphi^{\left( \sigma \right)} \right)}
\partial_\mu \varphi^{\left(\rho\right)} \partial_\nu \varphi^{\left(\rho\right)} \, .
\end{align}
}
The constraints in Eq.~(\ref{cnstrnt1}) are consistent with $\varphi^{\left(\rho\right)}=x^\rho$, which we now identify. It is then possible to rewrite Eq.~(\ref{Eqs3}) as
\textit{
\begin{align}
\label{Eqs4}
\Phi _{\left(\mu\nu\right)} \left(\varphi^{\left(\tau\right)}\right) = - 2 \mathcal{E}_{\mu\nu}
+ {\mathrm g}_{\mu\nu} \left\{ {\cal V} \left( \varphi^{\left(\rho\right)} \right) + \sum_{\rho=0,1,2,3} \lambda^{(\rho)} + {\mathrm g}^{\rho\sigma} \mathcal{E}_{\rho\sigma} \right\}
 - \frac{2 \lambda^{(\mu)}}{{\mathrm g}^{\mu\nu}\left( x^\sigma = \varphi^{\left( \sigma \right)} \right)} \delta_{\mu\nu} \, .
\end{align}
}
We pay particular attention to the solution for $\lambda^{(\rho)} =0$.
Once ${\mathrm g}_{\mu\nu}$ the arbitrary function ${\cal V} \left( \varphi^{\left(\rho\right)} = x^\rho \right)$ is provided. To achieve this, we determine $\Phi _{\left(\mu\nu\right)} \left(\varphi^{\left(\tau\right)}\right)$ as:
\textit{
\begin{align}
\label{Eqs5}
\Phi _{\left(\mu\nu\right)} \left(\varphi^{\left(\tau\right)}\right) = - 2 \mathcal{E}_{\mu\nu} \left( x^\sigma = \varphi^{\left( \sigma \right)} \right)
+ {\mathrm g}_{\mu\nu} \left( x^\sigma = \varphi^{\left( \sigma \right)} \right) \left\{ {\cal V} \left( \varphi^{\left(\rho\right)} \right)
+ {\mathrm g}^{\rho\sigma} \left( x^\sigma = \varphi^{\left( \sigma \right)} \right) \mathcal{E}_{\rho\sigma} \left( x^\sigma = \varphi^{\left( \sigma \right)} \right) \right\} \,,
\end{align}
}
Without loss of generality, we can  set ${\cal V} \left( \varphi^{\left(\rho\right)} \right)=0$ as it is arbitrary. The structure of the present study follows the pattern of a nonlinear sigma model, where  $\Phi _{\left(\mu \nu \right)} \left(\varphi^{\left(\tau\right)}\right)$ serves as the metric of the target space. The scalar field $\varphi \left(\mu \right)$ may be omitted if $\Phi _{\left(\mu \nu \right)}=0$ for a specific index $\mu$, $\mu=0\cdots 3$ and the remaining non-vanishing components of $\Phi _{\left(\rho \sigma \right)}$'s are not functions of $\varphi\left(\mu \right)$ using particular $\mu$.

If none of the eigenvalues of $\Phi _{\left(\mu\nu\right)} \left(\varphi^{\left(\tau\right)}\right)$ is negative and ${\Gamma}_\lambda$ is absent in Eq.~(\ref{acg1}), then  ghost degrees of freedom do not appear. We now investigate whether the constraints given in Eq.(\ref{cnstrnt1}) can eliminate potential ghosts arising from the term   ${\cal S} _\lambda$. To this end, we consider the perturbation introduced in Eq.(\ref{pert1}), expressed as:
\textit{
\begin{align}
\label{pert1B}
\varphi^{\left(\xi\right)}=x^\xi + \delta\varphi^{\left(\xi\right)} \, ,
\end{align}
}
where the perturbation of $\delta\varphi^\xi$ satisfies  Eq.~(\ref{cnstrnt1}) that yields:
\textrm{
\begin{align}
\label{pert2B}
0 = 2 {\mathrm g}^{\xi\nu}\left({ x^\sigma=\varphi^{(\sigma)}} \right) \partial_\nu \delta \varphi^{(\xi)}
 - \sum_\zeta \delta \varphi^{(\zeta)} \partial_\zeta {\mathrm g}^{\xi\xi}\left({ x^\sigma=\varphi^{(\sigma)}} \right) \, ,
\end{align}
}
where the above equations are not been summed over $\xi$.

In Eq.~(\ref{pert2B}), we may replace $\varphi^{(\sigma)}$ with $x^\sigma$ in the expressions  ${\mathrm g}^{\xi\nu}\left(x^\sigma=\varphi^{(\sigma)} \right)$ and ${\mathrm g}^{\xi\xi}\left(x^\sigma=\varphi^{(\sigma)} \right)$ provided that the perturbations from the background are properly taken into account.

If $\delta\varphi^{(\xi)}=0$ is imposed for a space-like coordinate $x^\xi$  in the limit $\left| x^\xi \right|\to \infty$, and also for a time-like coordinate,  $x^\xi$  then the condition  $\delta\varphi^{(\xi)}=0$ always holds. In this case, ghost degrees of freedom do not arise because $\delta\varphi^{(\xi)}=0$ does not propagate.

The scalar fields $\varphi^{\left(\xi\right)}$ are non-dynamical in the above discussion, as the perturbation is considered on a fixed background. Metric perturbations are excluded from the perturbation defined in Eq.~(\ref{pert1B}). However, since the scalar fields naturally couple to gravity, certain components of the metric may, in general, mix with the scalar field $\varphi^{\left(\xi\right)}$. As a result, Eq.(\ref{pert2B}) is modified to the following form: \[0 = 2 {\mathrm g}^{\xi\nu}\left(x^\sigma \right) \partial_\nu \delta \varphi^{(\xi)} - \sum_\zeta \delta \varphi^{(\zeta)} \partial_\zeta {\mathrm g}^{\xi\xi}\left(x^\sigma \right) - \delta {\mathrm g}^{\xi\xi} \left( x^\sigma \right).\]
Although the scalar fields $\varphi^{(\xi)}$ appear to have conjugate momenta, the imposed constraints prevent them from propagating. As a result, dynamical modes may not emerge, even when combinations of scalar modes are present. Furthermore, the scalar perturbation $\varphi^{(\xi)}$ does not mix with the tensor mode, despite the scalar fields
$\varphi^{(\xi)}$ being coupled to the  $+$-polarization. Although a more detailed analysis was undertaken, it did not lead to the emergence of any new propagating modes. This result could be further examined in future studies, particularly within the context of more general gravitational theories.

Mimetic gravity has been extended and explored through various approaches, as discussed in \cite{Takahashi:2017pje, Langlois:2018jdg, Ganz:2018mqi}, with several instabilities identified in the process. These results highlight the need for more thorough analyses to evaluate the model's consistency, which should be addressed in future studies.


As explained in Section \ref{f(T)}, it is not possible to obtain any spherically symmetric solution within the framework involving two scalar fields. Therefore, we now consider the functional form of  $f({\cal T})$ to be
\textrm{
\begin{align}
\label{fq}
f({\cal T})={\cal T}+\frac{\alpha}{2}{\cal T}^2\, .
\end{align}
}
We now demonstrate that it is possible to construct a four-scalar model within the framework of $f({\cal T})$ gravity, as defined in Eq.~(\ref{fq}), which admits a spherically symmetric spacetime.

When describing a stationary spherically symmetric spacetime, it is often more convenient to use Cartesian coordinates $\left(x, y, z\right) = \left(x^i\right)$ $\left(i = 1 \cdots 3\right)$ rather than $\left(r, \vartheta, \varphi\right)$. Now, through the use of the relation  \begin{align}& r = \sqrt{\sum_{i=1,2,3} \left(x^i\right)^2}, \,\, \mbox{ and } \, \,
{\mathit dr^2 + r^2 \left( d\vartheta^2 + \sin^2 \vartheta d\varphi^2 \right) = \sum_{i=1,2,3} \left( dx^i \right)^2},\nonumber\\
 & {\mathit rdr = \sum_{i=1,2,3} x^i dx^i}, \, \, \mbox{then  Eq.~(\ref{GBiv}),   with}\, \,\mu=\mu(r),\,  \mbox{and}\,\, \nu=\nu(r),\,\,  \mbox{can be reformulated as}
 \end{align}

\begin{align}
\label{ssmetric1}
{\mathit ds^2} =&\, {\mathit - \e^{2\mu(r)} dt^2 + \e^{2\nu(r)} dr^2 + r^2 \left( d\vartheta^2 + \sin^2 \vartheta d\varphi^2 \right)} \nonumber \\
=&\,{\mathit - \e^{2\mu\left(r\sqrt{\sum_{i=1,2,3} \left(x^i\right)^2}\right)} dt^2 + \left( \e^{2\nu\left(r=\sqrt{\sum_{i=1,2,3} \left(x^i\right)^2}\right)} -1 \right)
\frac{\left( \sum_{i=1,2,3} x^i dx^i \right)^2 }{\sum_{i=1,2,3} \left(x^i\right)^2}
+ \sum_{i=1,2,3} \left( dx^i \right)^2 }\nonumber \\
=&\, {\mathit{  \sum_{\mu,\nu=t, x, y, z} {\mathrm g}_{\mu \nu} dx^\mu dx^\nu}} \, ,
\end{align}
where,
\begin{align}
\label{ssmetric2_00}
{\mathrm g}_{tt} = -{\mathit  \e^{2\mu\left(r=\sqrt{\sum_{i=1,2,3} \left(x^i\right)^2}\right)}}\, , \quad
{\mathrm g}_{ij} = \left({\mathit  \e^{2\nu\left(r=\sqrt{\sum_{k=1,2,3} \left(x^k\right)^2}\right)} -1 } \right)
\frac{{\mathit  x^i x^j }}{{\mathit \sum_{k=1,2,3} \left(x^k\right)^2}} + {\mathit \delta_{ij}}\, ,
\end{align}

and
\textrm{
\begin{align}
\label{ssmetric2}
{\mathrm g}^{tt} = - {\mathit \e^{-2\mu\left(r\sqrt{\sum_{i=1,2,3} \left(x^i\right)^2}\right)}}\, , \quad
{\mathrm g}^{ij} = \left({\mathit  \e^{-2\nu\left(r=\sqrt{\sum_{k=1,2,3} \left(x^k\right)^2}\right)} -1} \right)
\frac{{\mathit x^i x^j }}{{\mathit \sum_{k=1,2,3} \left(x^k\right)^2}} + {\mathit \delta^{ij}}\, .
\end{align}
}
The tetrad field that reproduces the metric \eqref{ssmetric1} is given by:
\begin{align}\label{tetrad}
h^i{}_{\mu}=\left(
\begin{array}{cccc}
e^{\mu(r)} & 0 & 0 & 0 \\
\\
0 & 1+\frac{(e^{\nu(r)}-1)x^2}{r^2} & \frac{(e^{\nu(r)}-1)xy}{r^2} &\frac{(e^{\nu(r)}-1)xz}{r^2} \\
\\
0 &\frac{(e^{\nu(r)}-1)xy}{r^2}  &1+\frac{(e^{\nu(r)}-1)y^2}{r^2}  &\frac{(e^{\nu(r)}-1)yz}{r^2} \\
\\
0 &\frac{(e^{\nu(r)}-1)xz}{r^2} & \frac{(e^{\nu(r)}-1)yz}{r^2} & 1+\frac{(e^{\nu(r)}-1)z^2}{r^2} \\
\end{array}
\right).
\end{align}
The torsion scalar associated with the tetrad field \eqref{tetrad} takes the form:
\begin{align}
    {\cal T}=\frac{2(e^\nu(r)-1)(2e^{-2\nu(r)}\mu'r^2-re^{-\nu(r)}+re^{-2\nu(r)})}{r^3}
\end{align}
The detailed calculations of the components of the field equations of $f({\cal T})$ theory with four scalar fields, given by Eq.~\eqref{Eqs5} are very lengthy and not beneficial to list here.
 {
Utilizing Eq.~(\ref{ssmetric2}) and substituting Eq.~(\ref{fq}) into Eq.~(\ref{Qstatic}), we get:
\textrm{
\begin{align}
\label{df22}
&\Phi _{\left(t t\right)}=\frac {-2{e^{2 \mu  -4\nu  }}}{{r}^{4}} \left[  \left\{ {r}^{2} \left( {r}^{2}-2\beta \right) \mu''  +{r}^{2} \left( {r}^{2}-6 \beta \right) \mu'^{2}-r \left[ r \left( {r}^{2}-2\beta \right) \nu'  +12\beta-2{r}^{2} \right] \mu' -6\beta \right\} {e^{2\nu  }}\right.\nonumber\\
&\left.+ \left\{8{r}^{2} \left\{ r \left( {e^{\nu  }}-1 \right) \mu'  -1+2e^{\nu } \right\} \mu''  +4 \left(  \mu'  r+1 \right) {e^{3\nu  }}- {e^{4\nu  }}+ 4r^3\left( {e^{\nu  }}-1 \right)  \mu'^{3}+ 2\left\{ 2 \left( 3{r}^{3}-2{e^{\nu  }}{r}^{3} \right) \nu'  -5{r}^{2}+8{r}^{2}{e^{\nu  }} \right\} \mu'^{2}\right.\right.\nonumber\\
&\left.\left.+ 2\left\{ {r}^2 \left( 3-4{e^{\nu  }} \right) \nu'  +6{e^{\nu  }}r-2r \right\} \mu'  +4{ e^{\nu  }}-1 \right\}\beta \right]\,,\nonumber\\
&\Phi_{\left(i j\right)}= \frac{2\e^{-4\nu}}{r^6} \left\{ \left[ -2\left\{2\delta_{ij} +(4r\mu'+3)n_in_j \right\}r^4\beta \mu''-4 r^5 \mu'^3\beta n_i n_j+ 2\beta\left[ 6n_i n_j \nu'r-2\delta_{ij} +n_i n_j\right] r^4\mu'^2+r^3\mu'\left[\left(r^2-6\beta\right.\right.\right.\right.\nonumber\\
&\left.\left.\left.\left.+4\beta r\nu'\right)\delta_{i j}
 +\left(26\beta \nu'r-r^2+14\beta\right)n_i n_j\right] -r^3\nu'\left[\left(r^2-6\beta\right)\delta_{i j}-6\left(r^2+6\beta\right)n_i n_j\right] +r^2\left(r^2+6\beta\right)\delta_{ij}\right.\right.\nonumber\\
&\left.\left.+r^2\left(\beta-r^2\right)n_i n_j \right] {\e^{2\nu}}+{\e^{4\nu}}\left[\left\{\mu''r^2(r^2-10\beta)+r^2\mu'^2(r^2-6\beta)-r^2\nu'\mu'(r^2-10\beta)-2r(r^2-6\beta)+17\beta+r^2\right\}
r^2n_in_j\right.\right.\nonumber\\
&\left.\left.
+r^2(\beta-r^2)\delta_{i j}\right]+4\beta\left[r^2\mu''\left\{r^2n_in_j{\e^{3\nu}(r\mu'+5)}+r^2(\delta_{i j}-n_in_j)(2{\e^{2\nu}}-1)\right\}+{\e^{3\nu}}\left\{r^2n_in_j\left[r^3\mu'^3-2r^2\mu'^2\left[r\nu'-1\right]\right.\right.\right.\right.\nonumber\\
&\left.\left.\left.\left.-\mu'\left(3r+10r^2\nu'\right)
-6r\nu'-4\right]-r^2\delta_{i j}
\right\}+r^2n_in_j(r\mu'-1){\e^{5\nu}}-\frac{n_in_j}4{\e^{6\nu}}-r^2(n_in_j-\delta_{i j})\left\{r^2\mu'^2(3{\e^{\nu}}-2)\right.\right.\right.\nonumber\\
&\left.\left.\left.-r\mu'\left[(4{\e^{\nu}}-3)r\nu'-3{\e^{\nu}}+\frac{3}2\right]-3r\nu'\left({\e^{\nu}}-\frac{1}2\right)-{\e^{\nu}}+\frac{1}4\right\}\right]
\right\}\,,
\end{align}
}
where $\beta_1=\frac{\beta}{r^2}$ and $n_i=\frac{x_i} r$.  As we can see, the scalar fields $\Phi_{\left(\mu \nu\right)}$ are expressed in terms of the unknown functions $\mu$ and $\nu$. Therefore, by fixing an ansatz for $\mu$ and $\nu$,  we can derive the explicit forms of the scalar fields. In the next section, we present an application to a spherically symmetric spacetime to demonstrate how our construction operates.
{
\section{Spherically symmetric spacetime}\label{app}
Einstein introduced the principle of general covariance, which asserts that the mathematical formulation of physical laws should remain invariant under arbitrary coordinate transformations. In classical  GR, this principle is upheld by expressing physical laws in the form of tensor equations. However, in quantum gravity frameworks such as Loop Quantum Gravity, it remains uncertain whether the gravitational Hamiltonian's once quantum corrections are incorporated continues to respect gauge transformations and, by extension, preserves covariance \cite{Bojowald:2015zha}.

Recently, effective quantum gravity models that preserve covariance have been developed \cite{Zhang:2024khj}. In this section, we briefly review covariant, quantum-corrected black hole (BH) solutions obtained within the framework of effective quantum gravity. We then apply one such solution to our four-scalar model, derive the corresponding scalar fields, and analyze the thermodynamic properties of the system.

The metric describing the quantum-corrected BH spacetime takes the following form,
\begin{align}
\label{GBiva}
ds^2{\mathcal  = - \e^{2\mu (r)} dt^2 + \e^{2\nu (r)} dr^2 + r^2 \left( d\vartheta^2 + \sin^2\vartheta d\varphi^2 \right)}\,, \quad \mbox{where}\quad  \mu(r)=-\nu(r)=\ln\left[1-\frac{2M}{r}+\frac{\beta}{r^2}\left(1-\frac{2M}{r}\right)^2\right]^{1/2}\, ,
\end{align}
Here, $M$ and $\beta$ denote the ADM mass and the quantum correction parameter, respectively. Clearly, in the limit $\beta \to 0$, the quantum-corrected black hole reduces to the classical Schwarzschild solution. Applying the horizon condition $\nu(r) = 0$ yields $r_h = 2M$, indicating that the black hole horizon is unaffected by the quantum parameter $\beta$ and remains fixed regardless of its value.

The behavior of the metric given by Eq.\eqref{GBiva} is shown in Fig.\ref{Fig:Thr} \subref{fig:1a}. From this figure, it is evident that the presence of the dimensional parameter $\beta$ leads to the formation of two horizons, an inner and an outer one, unlike the Schwarzschild solution, which features a single horizon at $r_h$. Most notably, all three black hole horizons those corresponding to the two $\beta \neq 0$ cases and the one with $\beta = 0$, coincide at $r_h$, as clearly illustrated in Fig.~\ref{Fig:Thr} \subref{fig:1a}.

Next, we calculate the four scalars using the metric potentials $\mu(r)$ and $\nu(r)$ from Eq.\eqref{GBiva}. Due to the complexity and length of their explicit forms, we instead present their behavior graphically in Fig.\ref{Fig:Thr}. Specifically, the profiles of $\Phi_{(tt)}$, $\Phi_{(xx)}$, and $\Phi_{(xy)}$ are shown in Fig.~\ref{Fig:Thr} \subref{fig:1b}, \subref{fig:1c}, and \subref{fig:1d}, respectively. The remaining scalar fields $\Phi_{(yy)}$, $\Phi_{(zz)}$, $\Phi_{(xz)}$, and $\Phi_{(yz)}$ exhibit patterns similar to those of $\Phi_{(xx)}$ and $\Phi_{(xy)}$.
\begin{figure}
\centering
\subfigure[~The  behavior of the metric given by Eq. \eqref{GBiva}]{\label{fig:1a}\includegraphics[scale=0.25]{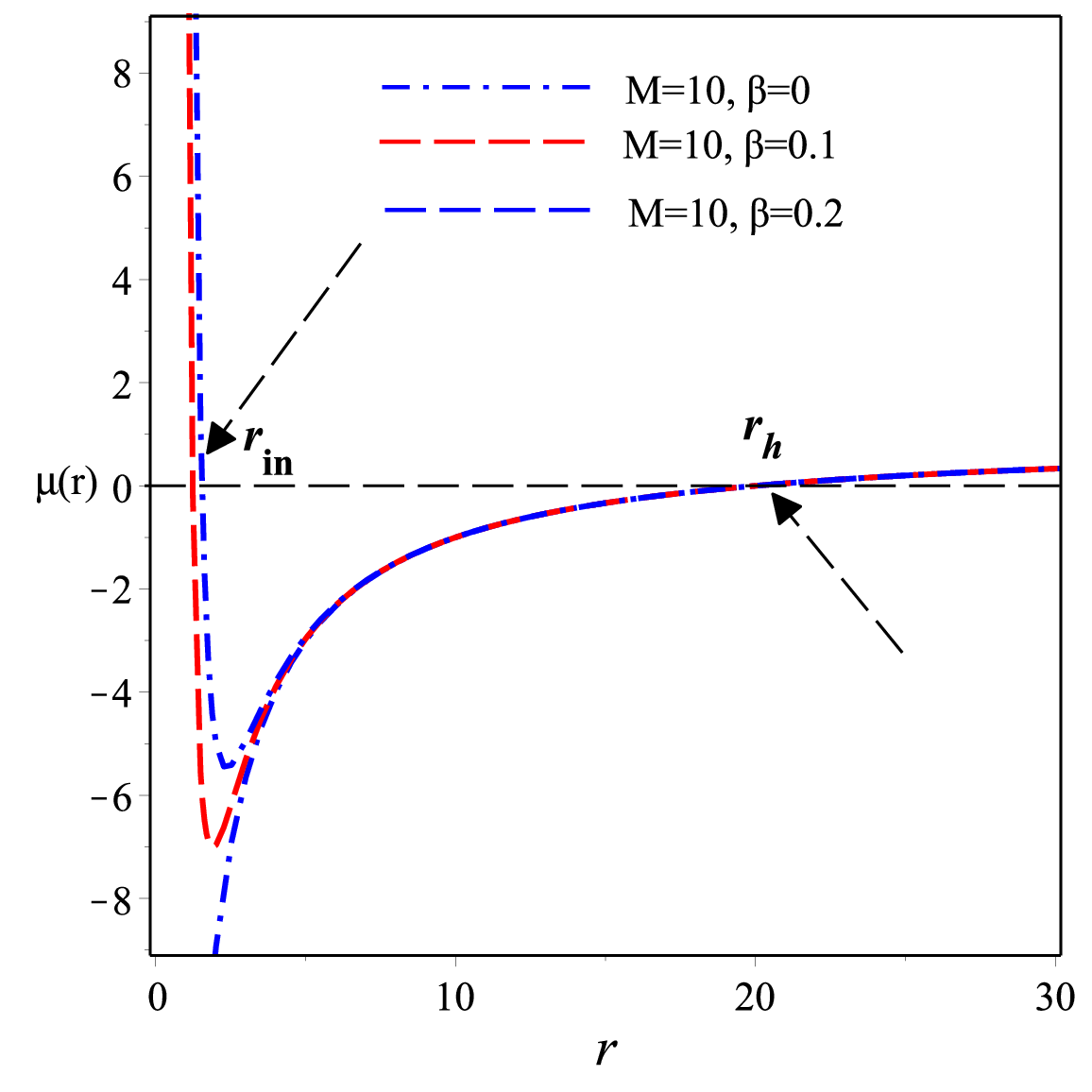}}\hspace{0.1cm}
\subfigure[~The  behavior of the scalar field  $\Phi_{(tt)}$  after using Eq. \eqref{GBiva}]{\label{fig:1b}\includegraphics[scale=0.25]{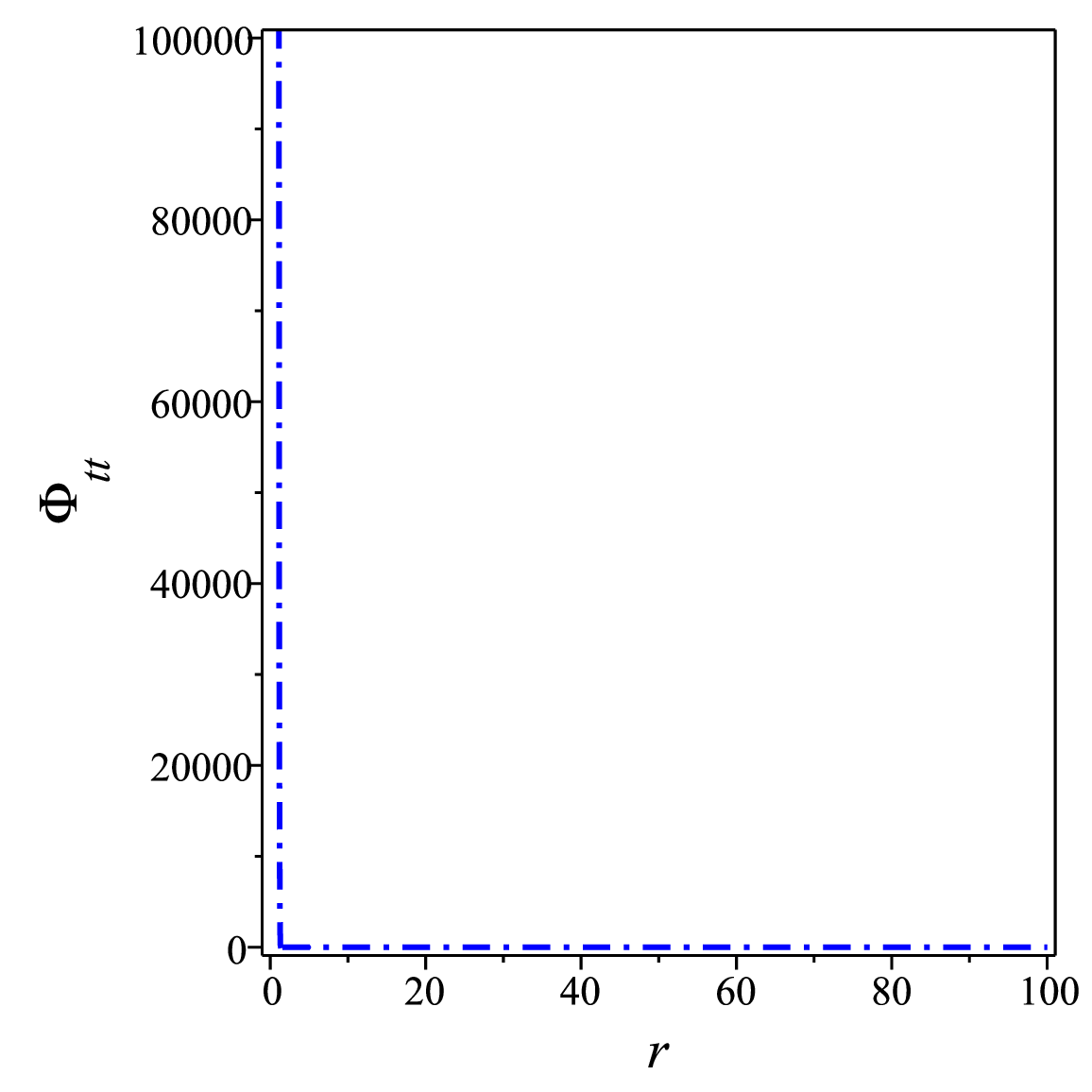}}\hspace{0.1cm}
\subfigure[~The  behavior of the scalar field  $\Phi_{(xx)}$  after using Eq. \eqref{GBiva}]{\label{fig:1c}\includegraphics[scale=0.25]{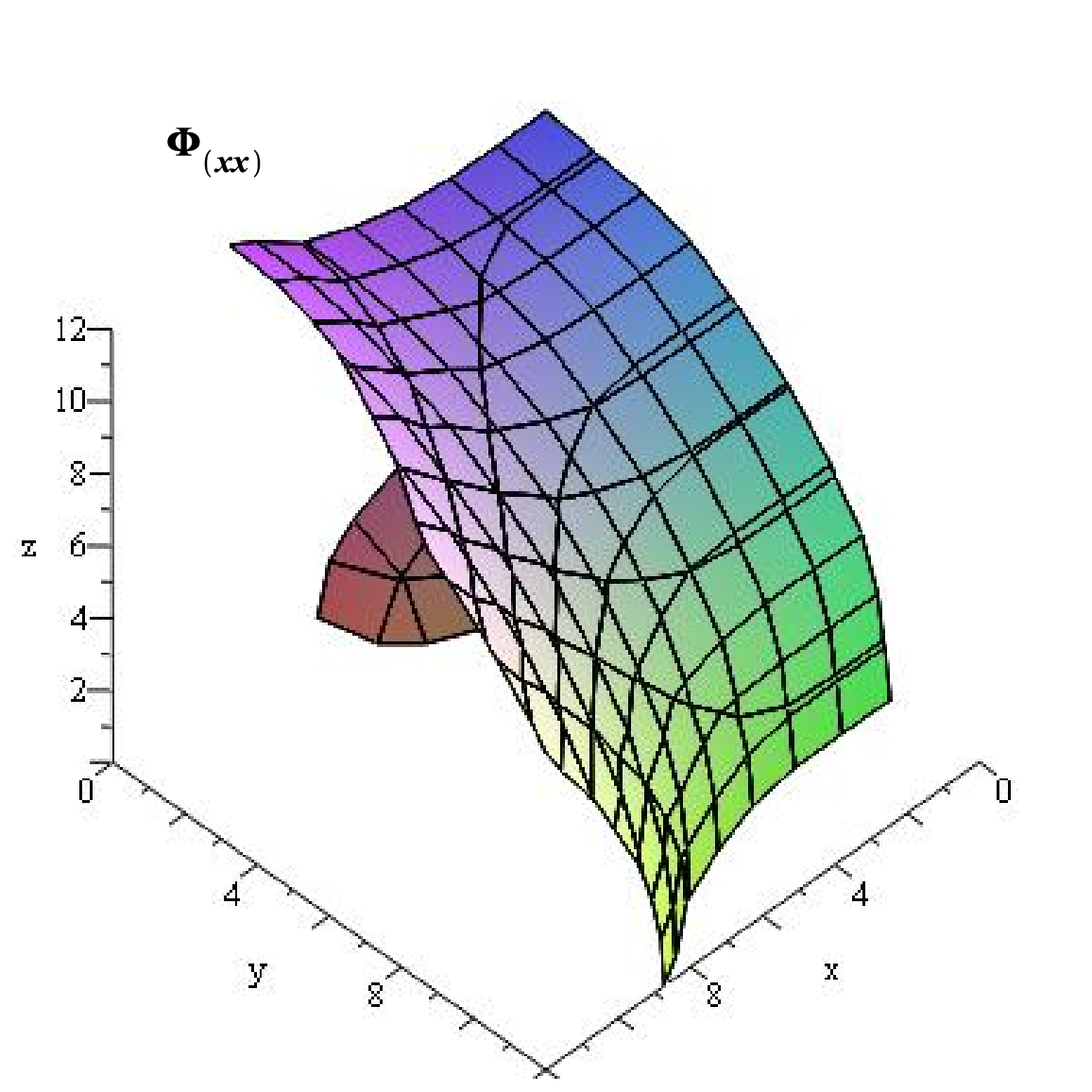}}\hspace{0.1cm}
\subfigure[~The  behavior of the scalar field  $\Phi_{(xy)}$  after using Eq. \eqref{GBiva}]{\label{fig:1d}\includegraphics[scale=0.25]{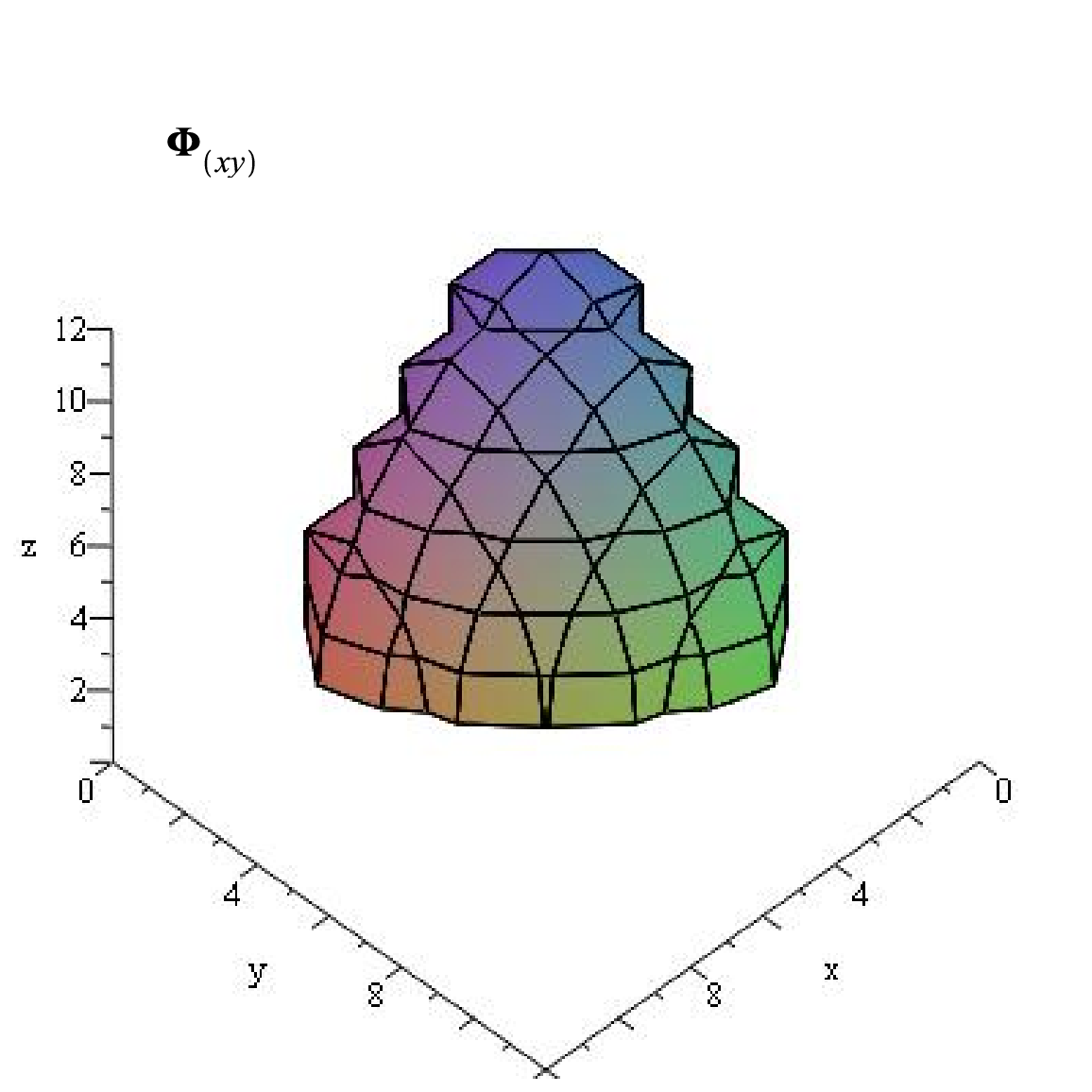}}
\hspace{0.1cm}
\subfigure[~The  behavior of the entropy given bu Eq. \eqref{ent11}]{\label{fig:1e}\includegraphics[scale=0.25]{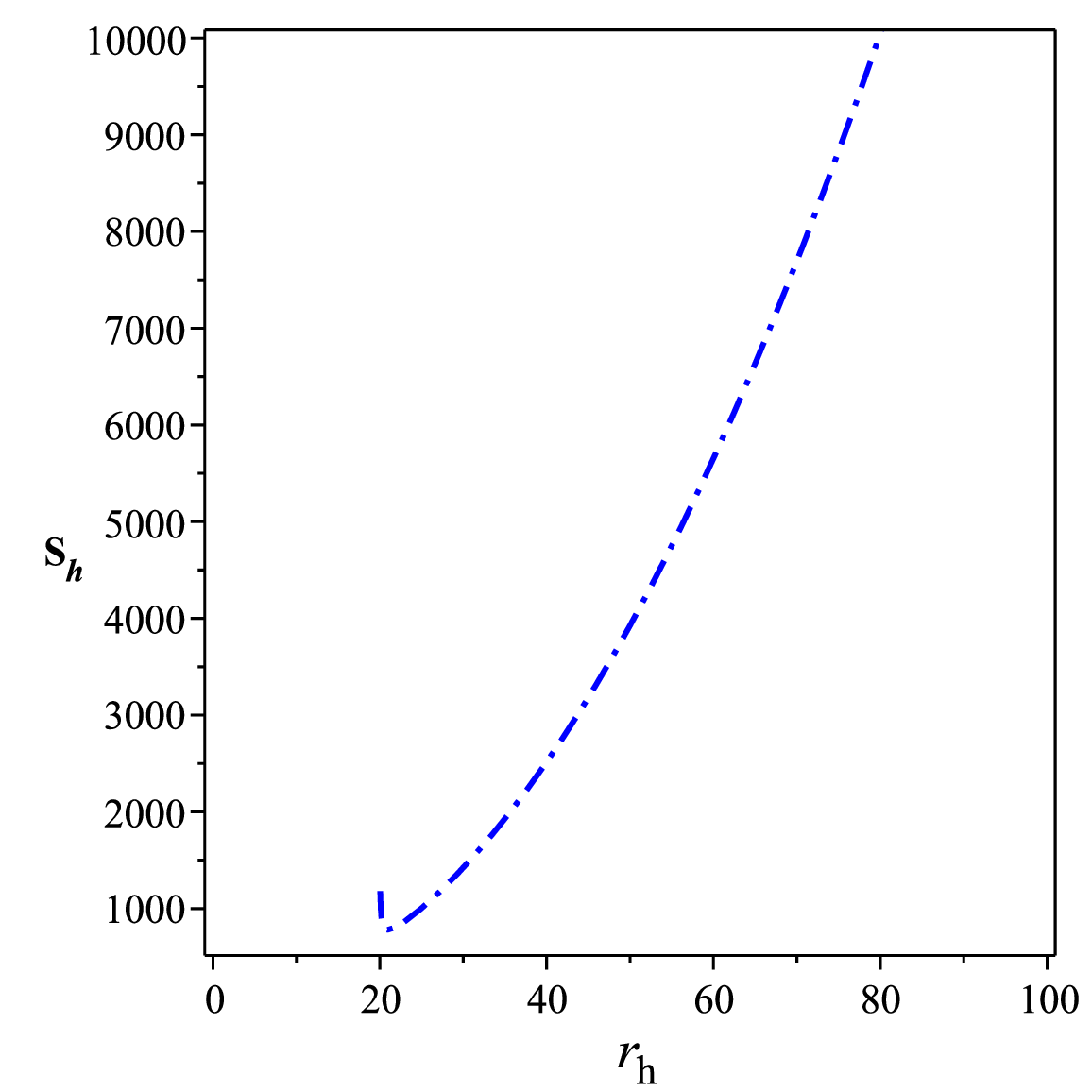}}
\hspace{0.1cm}
\subfigure[~The  behavior of the Hawking temperature]{\label{fig:1f}\includegraphics[scale=0.25]{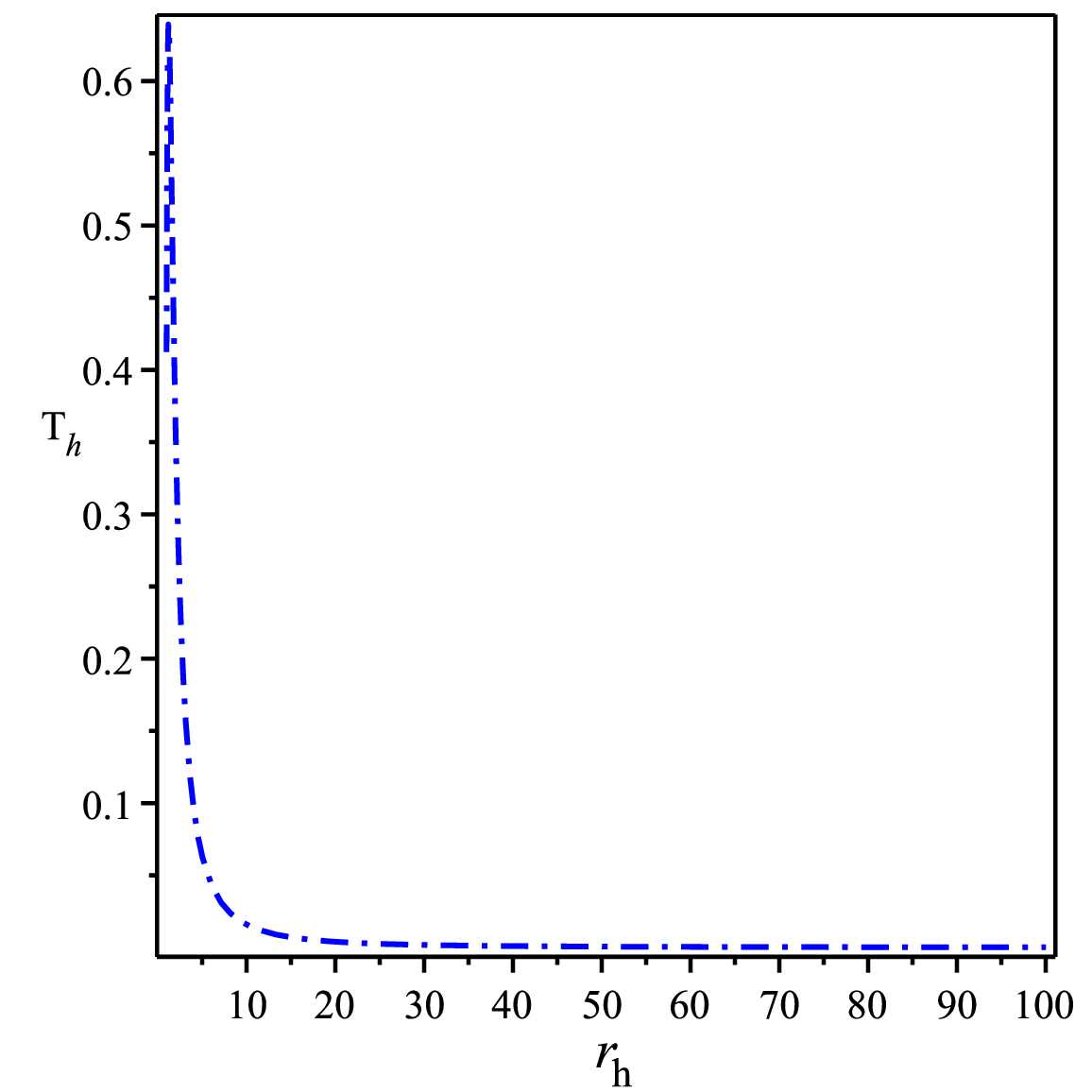}}
\hspace{0.1cm}
\subfigure[~The  behavior of the heat capacity]{\label{fig:1g}\includegraphics[scale=0.25]{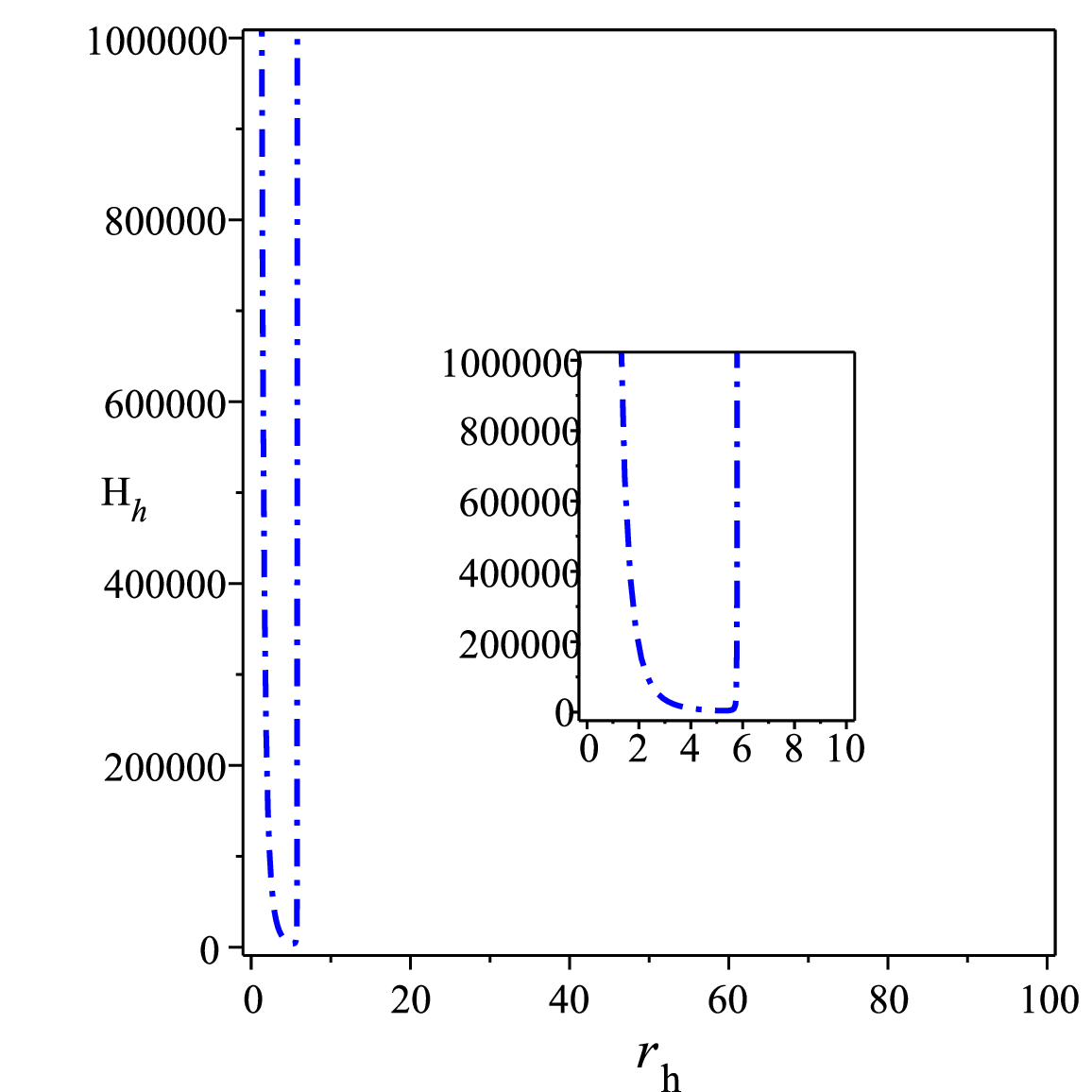}}
\hspace{0.1cm}
\caption{Fig.\subref{fig:1a} shows the general behavior of the metric described by Eq.\eqref{GBiva}. Fig.\subref{fig:1b} presents the profile of the scalar field $\Phi_{(tt)}$ derived using Eq.\eqref{GBiva}, while Fig.\subref{fig:1c} illustrates the behavior of the scalar field $\Phi_{(xx)}$. Fig.\subref{fig:1d} displays the pattern of the scalar field $\Phi_{(xy)}$. Fig.\subref{fig:1e} shows the Hawking temperature corresponding to the metric in Eq.~\ref {GBiva}, and Fig.~\subref{fig:1f} presents the behavior of the associated heat capacity. The model is characterized by the parameter values $\alpha = 10$, $\beta = 0.1$, and $M = 10$, which are used in the scalar field calculations. }
\label{Fig:Thr}
\end{figure}
Now let us calculate the thermodynamics of the metric \eqref{GBiva}.

In the context of ${f({\cal T})}$ gravity theory, the entropy is expressed as follows \cite{Nashed:2025qeu,Nashed:2024ush}:
\begin{equation}\label{ent11}
S(r_h)=\frac{1}{4}A(1+f_{\cal T}),
\end{equation}
where $A$ denotes the area and ${f_{\cal T}}$ represents the first derivative of ${f({\cal T})}$ with respect to  the torsion scalar. By substituting Eq. \eqref{GBiva} into Eq.~(\ref{ent11}), we obtain the entropy as follows:
\begin{align}\label{ent11}
&S(r_h)=\frac{1}{4}A(1+f_{\cal T})=\frac{\pi\,{r_h}^{2}}2 \left[ 1+\frac { 2\alpha\, {r_h}^{4}}{{r_h}^{6} \left(-2\,M{r_h}^{3}+\beta\,{r_h}^ {2}-4\,\beta\,r_hM+4\,\beta\,{M}^{2} \right) }\left\{ 4\,\Upsilon{r_h}^{3} \beta\,M-8\,\Upsilon{r_h}^{2}\beta\,{M}^{2}-2\,{r_h}^{8}-{r_h}^{6}\beta\right.\right.\nonumber\\
 &\left.\left.+{\beta}^{ 2}{r_h}^{4}+48\,{\beta}^{2}{M}^{4}+4\,M{r_h}^{7}+2\,\Upsilon{r_h}^{6}-2\,\Upsilon{r_h}^{ 5}M-2\,{r_h}^{5}\beta\,M+20\,{r_h}^{4}\beta\,{M}^{2}-24\,{M}^{3}{r_h}^{3} \beta-12\,{\beta}^{2}{r_h}^{3}M+48\,{\beta}^{2}{r_h}^{2}{M}^{2}\right.\right.\nonumber\\
 &\left.\left.-80\,{\beta }^{2}r_h{M}^{3} \right\} \right],
\end{align}
where $\Upsilon=\sqrt {{r_h}^{4}-2\,M{r_h}^{3}+ \beta\,{r_h}^{2}-4\,\beta\,r_hM+4\,\beta\,{M}^{2}}$.
The above equation suggests that the entropy is influenced by the higher-order torsion terms. The behavior of Eq. (\ref{ent11}) is depicted in Fig.~\ref{Fig:Thr} \subref{fig:1e}, demonstrating a positive value for entropy.

The Hawking temperature is typically defined in the following manner \cite{Sheykhi:2012zz,Sheykhi:2010zz,Hendi:2010gq,Sheykhi:2009pf}:
  \begin{equation}\label{temp11}
T_h = \frac{\kappa}{2\pi}=\frac{1}{4\pi}\left(\frac{dg_{tt}}{dr}\right)_{r\to r_h},
\end{equation}
where $\kappa$ represents the surface gravity of the BH. By applying Eq. \eqref{GBiva} to Eq. (\ref{temp11}), we derive the Hawking temperature as follows:
\begin{eqnarray}\label{T11}
&&T_h=\frac {M{r_h}^{3}-\beta\,{r_h}^{2}+6\,\beta\,r_hM-8\,\beta\,{M}^{2}}
{2\pi\,{r_h}^{5}}\,.
\end{eqnarray}
The characteristics of the Hawking temperature, as described by Eq. (\ref{temp11}), is depicted in Fig.~\ref{Fig:Thr} \subref{fig:1f}, demonstrating a positive value for temperature. This figure demonstrates that $T_h$ maintains a positive value throughout.

Additionally, examining the stability of the  BH  solution  is a critical topic that can be investigated both dynamically and via perturbative analysis \cite{Nashed:2003ee,Myung:2011we,Myung:2013oca}. To assess the thermodynamic stability of BHs, it is essential to derive the expression for the heat capacity, $H(r_h)$, evaluated at the event horizon. This quantity takes the form \cite{Nouicer:2007pu,hamblin:1999tk}:
\begin{equation}\label{heat-capacity11}
H_h=\frac{\partial M_h}{\partial T_h}=\frac{\partial M_h}{\partial r_h} \left(\frac{\partial T_h}{\partial r_h}\right)^{-1}\, .
\end{equation}
A  BH  is considered thermodynamically stable if its heat capacity, $H_h$, is positive; conversely, it is deemed unstable if $H_h$ is negative. By substituting equations (\ref{ent11}) and (\ref{T11}) into equation (\ref{heat-capacity11}), we derive the heat capacity as follows:
\begin{align}\label{heat-cap1a11}
&{H_+}_{{}_{{}_{\tiny Eq.\eqref{GBiva}}}}= \left[ \pi \left( 192{M}^{4}{\beta}^{2}\Upsilon\alpha-64\alpha{r_h}^{ 2}{\beta}^{2}{M}^{4}-288{M}^{3}{\beta}^{2}\Upsilon\alpha r_h+96\alpha{ r_h}^{3}{\beta}^{2}{M}^{3}-96{M}^{3}\beta\Upsilon\alpha{r_h}^{3}+64 \alpha{r_h}^{5}\beta{M}^{3}\right.\right.\nonumber\\
&\left.\left.-48\alpha{r_h}^{4}{\beta}^{2}{M}^{2}+ 152{M}^{2}{\beta}^{2}\Upsilon\alpha{r_h}^{2}-72\alpha{r_h}^{6}\beta {M}^{2}+96{M}^{2}\beta\Upsilon\alpha{r_h}^{4}-4{M}^{2}\beta\Upsilon{r_h}^ {6}+2\alpha{r_h}^{8}{M}^{2}-32M\Upsilon\alpha{\beta}^{2}{r_h}^{3}\right.\right.\nonumber\\
&\left.\left.+8 \alpha{r_h}^{5}{\beta}^{2}M-28M\beta\Upsilon\alpha{r_h}^{5}+24 \alpha{r_h}^{7}\beta M+4M\beta\Upsilon{r_h}^{7}+2M\Upsilon{r_h}^{9}+2\Upsilon \alpha{\beta}^{2}{r_h}^{4}-\beta{r_h}^{8}\Upsilon+2\Upsilon\alpha{r_h}^{6} \beta-2\alpha{r_h}^{8}\beta-\Upsilon{r_h}^{10} \right) \right.\nonumber\\
&\left. \left( \beta{r_h}^{2}-M{r_h}^{3}-6\beta r_hM+8\beta{M}^{2} \right) \right]\left\{ \left( 3\beta{r_h}^{2}-2 M{r_h}^{3}-24\beta r_hM+40\beta{M}^{2} \right) \left( {r_h}^{4}-2M{r_h}^{3}+\beta{r_h}^{2}-4\beta r_hM \right.\right.\nonumber\\
&\left.\left.+4\beta{M}^ {2} \right) ^{3/2}{r_h}^{4}\right\}^{-1}.
\end{align}
Equation (\ref{heat-cap1a11}) indicates that $H_+$ does not exhibit any local divergence. The behavior of the heat capacity is illustrated in Fig.~\ref{Fig:Thr} \subref{fig:1g}, demonstrating a positive trend, which implies that the BH described by Eq. (\ref{GBiva}) is stable.

Thus, we have successfully derived a spherically symmetric spacetime within the four-scalar model, which could not be obtained in either the mimetic theory or the two-scalar model. We summarize our results in the following table.\\
\\
\begin{table}
\caption{Comparison between different theories and their ability to reproduce spherically symmetric spacetime}
 \begin{tabular}
 {|c|c|c|c|}
     \hline
     Theory& GR with two scalar fields & $f(T)$ with two scalar fields & $f(T)$ four scalar fields \\ \hline
    spherical symmetry spacetime &can be derived  & cannot be derived&can be derived   \\ \hline
     \end{tabular}
     \end{table}
 }
}
\section{Summary and discussion}
\label{summaryanddiscussion}
In this work, we examined the realization of spherically symmetric solutions in the framework of modified teleparallel gravity, specifically within
$f({\cal T})$ theory. It is well known that non-linear
$f({\cal T})$  theories impose stringent conditions on the torsion scalar
${\cal T}$, often requiring it to be constant to admit spherically symmetric solutions. This restricts such models to be dynamically equivalent to the teleparallel equivalent of general relativity (TEGR), effectively limiting their physical viability for describing general spacetimes.

We began by reviewing the two-scalar field model previously used in Einstein gravity to reproduce arbitrary spherically symmetric spacetimes. However, in the context of $f({\cal T})$ gravity, we showed that this model either requires
$f({\cal T})$ to be linear (i.e., TEGR) or demands a constant torsion scalar, both of which hinder the development of genuinely modified gravitational dynamics.

To overcome this, we proposed a novel four-scalar field model. This extension allows the construction of arbitrary spherically symmetric geometries within a general $f({\cal T})$ framework, including non-linear cases. By introducing appropriate Lagrange multipliers, we ensured the elimination of ghost degrees of freedom, maintaining the model's physical consistency.

We applied this formulation to demonstrate that the Schwarzschild solution can be recovered in $f({\cal T})$ gravity with a quadratic correction, i.e.,
$f({\cal T})= {\cal T}+\frac{\alpha}{2}{\cal T}$. This marks a significant advancement, as such a solution was previously inaccessible in standard
$f({\cal T})$ approaches due to the non-constancy of the torsion scalar.

{ The core motivation behind our four-scalar model lies in addressing a fundamental structural issue in
$f({\cal T})$ gravity. The field equations of non-linear $f({\cal T})$ gravity impose a highly restrictive condition:
$$\frac{d{\cal T}}{dr} \frac{d^2f({\cal T})}{d {\cal T}^2}=0.$$. The above condition implies that, unless $f({\cal T})$ is linear (i.e.,
$f({\cal T})={\cal T}+const.)$, one cannot derive spherically symmetric solutions. In other words, either
${\cal T}=const.$, corresponding to TEGR, or the form of
$f({\cal T})$ collapses to linearity, again reducing to TEGR. This severely limits the predictive power of non-linear $f({\cal T})$ models in physically relevant settings.  Although the two-scalar model is employed, one cannot derive spherically symmetric spacetime because of the non-vanishing of Eq.~(\ref{rtheta}) which contradicts the constrains presented in Eq.~\eqref{TSBH2B} that is necessary for deriving spherically symmetric solutions.

The issue of spherically symmetric solution in $f({\cal T})$ has been well documented in the literature. Some researchers have attempted to bypass it by abandoning spherically symmetric configurations and used cylindrical spacetimes (e.g., \cite{Capozziello:2019uvk,Nashed:2017fnd,Awad:2017tyz}), while others have explored perturbative approaches (e.g.,\cite{Bahamonde:2019zea,DeBenedictis:2022sja}).
{ In this study, we successfully derived a four-scalar model. By specifying a particular form of a spherically symmetric spacetime, we obtained the corresponding four scalar fields and demonstrated that all of them exhibit positive behavior, ensuring that our construction is free from ghost instabilities. Furthermore, we computed the thermodynamic quantities associated with this spacetime and analyzed its stability by evaluating the heat capacity, which was found to be positive, indicating thermodynamic stability.}

 As far as we know that within $f({\cal T})$ theory (even in its special case like the quadratic or cubic form) there is no study has been done to construct a stellar model. The main reason for not constructing a stellar model within this theory is the junction condition, i.e., how we make a junction condition between the interior solution and the exterior solution which is not known. Through the use of the four scalar model we can now construct a stellar model (The study of stellar model is out the scope of the present research) where we can make a junction condition between its interior and its exterior which is the Schwarzschild solution.

 { Finally, to conclude our study, we emphasize that our four-scalar model differs significantly from multi-scalar and mimetic models in several key aspects. First, we do not assume any conformal transformation of the metric, as is commonly done in mimetic gravity. Second, most multi-scalar and mimetic models are developed within a cosmological context to address challenges specific to that domain. To the best of our knowledge, these theories have not yet been applied in the astrophysical context; specifically, no stellar models have been constructed within such frameworks. In contrast, our four-scalar model allows for such applications, opening new avenues for exploring astrophysical phenomena.}


\section{Acknowledgments}
{This work was supported and funded by the Deanship of Scientific Research at Imam Mohammad Ibn Saud Islamic University (IMSIU) (grant number IMSIU DDRSP2502).}

%

\end{document}